\documentclass[
superscriptaddress,
amsmath,amssymb,
aps, 
prb,
twocolumn, 
floatfix, 
longbibliography
]{revtex4-2}
\usepackage{graphicx}
\usepackage{siunitx}
\usepackage{amsmath}
\usepackage{hyperref}
\usepackage{cleveref}
\usepackage{float}

\newcommand{\vect}[1]{\boldsymbol{\mathbf{#1}}}
\newcommand{\op}[1]{{\hat{#1}}}
\newcommand{\hc}[1]{{\hat{#1}}^{\dagger}}

\crefname{figure}{Fig.}{Figs.}
\Crefname{figure}{Figure}{Figures}

\crefname{table}{Tab.}{Tabs.}
\Crefname{table}{Table}{Tables}

\crefname{appendix}{Sec.}{Secs.}
\Crefname{appendix}{Appendix}{Appendices}
\crefname{section}{Sec.}{Secs.}
\Crefname{section}{Section}{Sections}
\crefname{equation}{Eq.}{Eqs.}
\Crefname{equation}{Equation}{Equations}

\crefmultiformat{equation}
  {Eqs.~(#2#1#3)}            
  { and~(#2#1#3)}            
  {, (#2#1#3)}               
  { and~(#2#1#3)}            

\crefrangelabelformat{equation}{(#3#1--#2#4)}

\begin{document}

\title{Model of the magnon Kerr effect in a highly anisotropic ferromagnet}

\author{Davit Petrosyan}
\email[]{davit.petrosyan@mat.ethz.ch}

\affiliation{Department of Materials, ETH Zurich, CH-8093 Zurich, Switzerland}
\author{Hiroki Matsumoto}

\affiliation{Department of Materials, ETH Zurich, CH-8093 Zurich, Switzerland}
\affiliation{Institute for Chemical Research, Kyoto University, 6110011 Uji, Japan}
\author{Hanchen Wang}

\affiliation{Department of Materials, ETH Zurich, CH-8093 Zurich, Switzerland}
\author{Richard Schlitz}

\affiliation{Department of Physics, University of Konstanz, 78457 Konstanz, Germany}

\author{Pietro Gambardella}

\affiliation{Department of Materials, ETH Zurich, CH-8093 Zurich, Switzerland}

\author{William Legrand}
\email[]{william.legrand@neel.cnrs.fr}

\affiliation{Department of Materials, ETH Zurich, CH-8093 Zurich, Switzerland}
\affiliation{Université Grenoble Alpes, CNRS, Institut Néel, 38042 Grenoble, France}

\begin{abstract}
The magnon Kerr effect is a nonlinear phenomenon that occurs universally in any ferromagnet with finite magnetic anisotropy. It manifests itself as a magnon population-dependent change in the magnon dispersion relation. Here, we derive the Hamiltonian for the magnon Kerr effect for a thin film with uniaxial anisotropy. We consider the external field, parallel, perpendicular, and at an intermediate angle to the anisotropy axis, and determine the Kerr coefficient in each case. We show how the nonlinearity scales for different magnetic materials as a function of magnon frequency and magnetic sample volume. Moreover, we derive the equations of motion for a hybrid cavity--magnon system with such thin-film magnon Kerr nonlinearity, to obtain populations of photons and magnons, and scattering matrix terms for the microwave transmission of the cavity, relevant for experimental observations.
\end{abstract}

\maketitle

\section{Introduction}
Nonlinear effects have become a cornerstone of magnonics, aiming to utilize the nonlinear dynamics of magnons for the exploration of new physical phenomena and devices \cite{Kruglyak_2010,Serga_2010, pirro2021advances, ZARERAMESHTI20221}. Nonlinear magnon--magnon interactions commonly appear for significant magnon populations, and can be used to generate entanglement with other bosonic systems \cite{PhysRevLett.121.203601,PhysRevLett.121.137203,PhysRevB.99.134426,PhysRevResearch.2.013154}. Examples of nonlinear magnonic phenomena include three-magnon mixing \cite{SUHL1957209, Kurebayashi2011, PhysRevLett.103.157202, PhysRevLett.130.046703, PhysRevLett.85.2184, PhysRevB.67.104402, PhysRevB.79.144428, PhysRevLett.103.157202, jnpb-2mxx}, parametric pumping of magnons \cite{PhysRevB.86.134420, BRACHER20171, 10.1063/5.0038946}, Bose-Einstein condensation of magnons \cite{PhysRevLett.99.037205, serga_boseeinstein_2014}, magnon self Kerr effect \cite{PhysRevB.94.224410}, and cross Kerr \cite{PhysRevLett.129.123601, Shen2025} effects. 

The self Kerr effect of magnons, here on referred to as the magnon Kerr effect (MKE), appears in any ferromagnetic system with finite magnetic anisotropy, making it a universal phenomenon in magnonics. It stems from the quadratic energy terms associated to magnetic anisotropy, and can be modeled as a Duffing-type nonlinearity in the oscillator equations \cite{PhysRevB.101.054402}. It describes the action of the anisotropy field on the magnetization dynamics as the system is populated by more magnons, causing a magnon-population-dependent change in the dispersion relation. This nonlinearity scales as the effective magnetic anisotropy. The MKE was experimentally demonstrated using yttrium iron garnet (YIG) spheres at cryogenic temperatures, driving with large power their spatially uniform Kittel mode in a hybrid microwave cavity--magnon system \cite{PhysRevB.94.224410}. The MKE also gives rise to bistability in magnons \cite{PhysRevLett.120.057202}, and has been proposed to generate squeezed states \cite{PhysRevA.111.013708, https://doi.org/10.1002/qute.202400654}, nonreciprocal magnon interactions \cite{PhysRevA.110.043704, PhysRevApplied.12.034001}, magnon blockade \cite{BHATT2025172275}, and magnon--magnon entanglement \cite{PhysRevResearch.1.023021}. The universality of the MKE and its compatibility with other bosonic platforms \cite{Awschalom2021,YUAN20221} make it an important phenomenon for nonlinear cavity magnonics, provided the MKE can be strengthened in selected magnetic systems. 

Magnetic thin films are particularly suitable for reinforcing these nonlinear behaviors. Recent experiments show that the MKE is strongly enhanced in thin films compared to bulk spheres, owing to shape anisotropy \cite{x4f8-pfj2}. Additionally, magnetic anisotropy can be further controlled by tailoring the material and its interfacial properties at the nanoscale through epitaxial growth \cite{PhysRevB.89.134404, Soumah2018}, which enables tuning of the magnon dispersion and control of magnon nonlinearities for higher-order standing spin-wave modes. The magnon nonlinearity can also be considerably enhanced by patterning, which further reduces the magnetic volume and provides additional interaction terms due to finite-size effects. These rich prospects motivate the investigation of the MKE in thin films~\cite{x4f8-pfj2}.

In this work, we derive the Hamiltonian of the MKE in a ferromagnetic thin film. For in-plane (IP), out-of-plane (OOP), and intermediate-angle magnetic configurations, we calculate the magnon Kerr coefficient $\mathcal{K}$, such that each additional magnon in the system shifts the magnon frequency by $2\mathcal{K}$. In particular when the external field is not OOP, the anisotropy field and the external field are not collinear. This corresponds to a highly anisotropic energy landscape, which differs from previous models that have been established for spheres \cite{PhysRevB.94.224410}. We then predict the anharmonicity considering actual material parameters for both magnetic insulators and metallic ferromagnets. We derive a strong enhancement of the MKE due to the shape anisotropy arising from the thin-film geometry, which is absent in spheres, in agreement with recent measurements~\cite{x4f8-pfj2}. Our model highlights the role of the ellipticity of the magnetic precession in the MKE of thin films with both IP and intermediate-angle magnetic configurations, and the power-dependent dynamical reorientation effect affecting the nonlinearity at intermediate angles. Accounting for all these factors, we then derive the equations of motion (EOM) and simulate magnon and photon populations in a linearly coupled magnon--photon hybrid resonator, the most common experimental system in cavity magnonics. The cavity acts as a tool to drive and probe the magnonic system, hence, we extend our theory to the case of a hybrid cavity--magnon system, in order to provide insight into common experimental conditions. We relate the Kerr coefficient and the drive power to the populations reached by the magnon and photon components of the hybrid modes. Finally, we simulate the corresponding microwave transmission spectra in a two-port configuration, whose evolution with increasing input microwave power links the Kerr coefficient to experimentally accessible quantities.

In \cref{sec:spheres_vs_films}, we compare the magnetic anisotropy energy, which gives rise to the MKE, in YIG spheres and thin films. Then, in  \cref{sec:hamiltonian}, we derive the MKE Hamiltonian for a ferromagnetic thin film magnetized OOP, IP, and at intermediate angles, starting from the classical energies of the magnetic system. In \cref{sec:eom}, we derive the EOM for the thin-film magnons strongly coupled to a two-port cavity resonator, enabling the determination of the magnon Kerr coefficient from experimental observations. Finally, in \cref{sec:results}, we simulate the behaviors expected in experiments considering different relevant magnetic materials, and we discuss the prospects for future implementations.

\section{Magnetic anisotropy energy landscape in spherical and thin-film systems}
\label{sec:spheres_vs_films}
\begin{figure}[t]
\includegraphics[width=\linewidth]{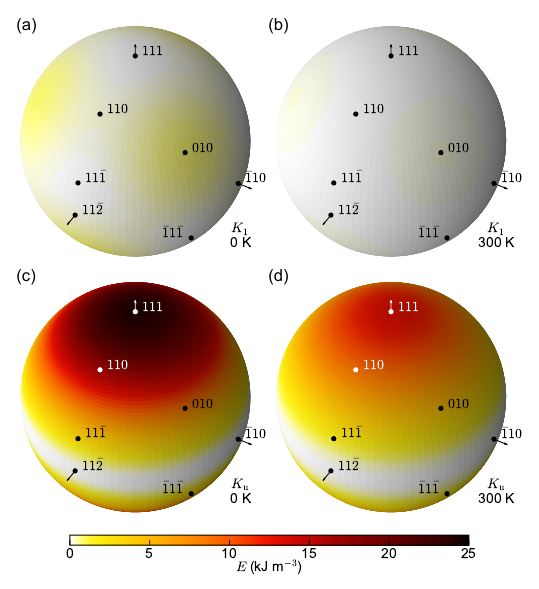}%
\caption{Magnetocrystalline (a,b) and magnetostatic shape (c,d) anisotropy energies in YIG at \SI{0}{K} (a,c) and at \SI{300}{K} (b,d), respectively, along its crystallographic directions. The thin film limit is used for (c) and (d) with OOP direction along the [111] axis.}
\label{fig:fig_spheres}
\end{figure}

To illustrate the differences occurring between a spherical sample of a magnetically ordered material, for which the MKE originates from magnetocrystalline anisotropy, and a thin film, for which the MKE might originate from strongly uniaxial or shape anisotropy, we consider the example of YIG. It is indeed the prototypical material used in magnonics, owing to its ultralow magnetic damping \cite{pirro2021advances, Serga_2010, Kruglyak_2010}. In particular, YIG spheres exhibit the longest magnon lifetimes that have been identified \cite{PhysRevLett.3.32, 10.1063/5.0306423, doi:10.1126/sciadv.aee2344} and are widely used for inducing nonlinear magnonic phenomena \cite{ZARERAMESHTI20221}, including the MKE \cite{PhysRevB.94.224410}. The magnetic anisotropy energy densities are given, for the first-order cubic anisotropy, by $E_{\rm{mc}}=K_1(\alpha_1^2\alpha_2^2+\alpha_2^2\alpha_3^2+\alpha_3^2\alpha_1^2)$, where $\alpha_{1,2,3}=\vect{m}\cdot\vect{e}_{1,2,3}$ with $\vect{m}$ the orientation vector of the magnetization and denoting by $\vect{e}_{1,2,3}$ the [100], [010], and [001] cubic crystallographic directions; for the uniaxial anisotropy, by $E_{\rm{u}}=-K_{\rm{u}}(\alpha_z^2)$, where $\alpha_{z}=\vect{m}\cdot\vect{e}_{z}$ with $\vect{e}_{z}$ along the out-of-plane direction $z$, and $K_{\rm{u}}=-\mu_0M_{\rm{s}}^2/2$ for a purely magnetostatic contribution. The energy landscape of the magnetocrystalline anisotropy in YIG is shown at \SI{0}{K} ($K_1=-\SI{2.48}{kJ.m^{-3}}$ \cite{Hansen1974a}) in \cref{fig:fig_spheres}(a) and at \SI{300}{K} ($K_1=-\SI{0.6}{kJ.m^{-3}}$ \cite{Hansen1974a}) in \cref{fig:fig_spheres}(b). For a thin film, this corresponds to a (111)-oriented growth. Comparing with the magnetostatic shape anisotropy at \SI{0}{K} ($M_{\rm{s}}=\SI{196}{kA.m^{-1}}$ \cite{Hansen1974a}, $K_{\rm{u}}\approx-\SI{24}{kJ.m^{-3}}$) in \cref{fig:fig_spheres}(c) and at \SI{300}{K} ($M_{\rm{s}}=\SI{143}{kA.m^{-1}}$ \cite{Hansen1974a}, $K_{\rm{u}}\approx-\SI{13}{kJ.m^{-3}}$) in \cref{fig:fig_spheres}(d), the magnetocrystalline anisotropy is one to two orders of magnitude lower. Combined with the lower bound on the volume of the spheres (diameter $>\SI{0.2}{mm}$), the strength of the MKE in bulk remains ultimately limited. The magnetostatic shape anisotropy for YIG films is considerable both at low and at room temperatures, and factoring in the ability to lower the volume by varying the thickness of the film \cite{x4f8-pfj2} or by patterning \cite{zw18-26nw}, the strength of the MKE can be controlled much more extensively.

\section{Theoretical Model}
\label{sec:theory}

\subsection{Magnetic Hamiltonian}\label{sec:hamiltonian}
We first derive the MKE Hamiltonians for a thin film magnetized OOP, IP, and at an intermediate angle. They are obtained from the classical energies of the magnetic system, taking into account Zeeman and anisotropy terms defined by 
\begin{equation}\label{eq:Hamiltonian_start}
    H_\mathrm{mag} = -\mu_0\int_{V_{\rm{m}}}\vect{M}\cdot\vect{H}_{\rm{ext}}\ \mathrm{d}V - \frac{\mu_0}{2}\int_{V_{\rm{m}}}\vect{M}\cdot\vect{H}_{\rm{a}}\ \mathrm{d}V,
\end{equation}
with magnetization vector $\vect{M}$, external field $\vect{H}_{\rm{ext}}$, and anisotropy field $\vect{H}_{\rm{a}}=2K_{ \rm{a}}(\vect{M}\cdot\vect{e}_{\rm{a}})/(\mu_0M_{\rm{s}}^2)\ \vect{e}_{\rm{a}}$, containing the anisotropy energy density $K_{\rm{a}}$, the saturation magnetization $M_{\rm{s}}$, and the magnetic anisotropy direction $\vect{e}_{\rm{a}}$. For $K_{\rm{a}}>0$, the anisotropy is uniaxial along $\vect{e}_{\rm{a}}$, while for $K_{\rm{a}}<0$, the anisotropy is easy-plane in a plane normal to $\vect{e}_{\rm{a}}$. The components of the total spin operator fulfill $\gamma{}S_{x,y,z}=M_{x,y,z}V_{\rm{m}}$, with $\gamma$ the gyromagnetic ratio and $V_\mathrm{m}$ the volume of the magnetic sample. They will then be mapped to the bosonic creation and annihilation operators following a Holstein-Primakoff transformation \cite{Holstein1940}.

So far, MKE models have been derived for a bulk magnetic sphere \cite{PhysRevB.94.224410}, where the anharmonicity arises from the magnetocrystalline anisotropy with direction aligned with the external magnetic field. With only $\op{S}_z$ and $\op{S}_z^2$ terms in the Hamiltonian, a factorized form is easily obtained. The present situation for IP and intermediate fields is conceptually different, because the external field couples to $\op{S}_z$ spin operator, while the anisotropy term couples to $\op{S}_x^2$ spin operator, and these spin operators do not commute with each other. Comparing OOP and IP magnetic configurations, we will evidence the opposite sign for their magnon Kerr coefficient $\mathcal{K}_{\rm{OOP/IP}}$, before extending to the angular dependence of the magnon Kerr coefficient.

\subsubsection{Out-of-plane field}
First, we consider an OOP external field, and a uniaxial anisotropy $K_{ \rm{u}}=K_{\rm{a}}$ with $\vect{e}_{\rm{a}}$ along $z$, resulting from the sum of any anisotropy mechanisms along the film normal and of the effective shape anisotropy due to magnetostatic terms. When $\vect{H}_{\rm{ext}}$ and $\vect{e}_{\rm{a}}$ are aligned, the OOP magnetic Hamiltonian $\op{H}_\mathrm{OOP}$ is therefore
\begin{equation}\label{eq:Hamiltonian_oop}
    \op{H}_\mathrm{OOP} = -\gamma B_\mathrm{ext}\op{S}_z-\frac{\gamma^2K_{\rm{u}}}{M_{\rm{s}}^2V_{\rm{m}}}\op{S}_z^2,
\end{equation}
where $B_\mathrm{ext}=\mu_0 H_\mathrm{ext}$. For a number of spin excitations small compared to the total number of spins $N$, the total spin operator is well described in terms of magnon creation and annihilation operators, relying on the Holstein-Primakoff transformation \cite{Holstein1940}. New operators are introduced and linearized
\begin{align}
    \label{eq:HP_Splus}
    \op{S}_{+}&=\op{S}_{x}+i\op{S}_{y}=\left(\sqrt{2S-\hc{b}\op{b}}\right)\op{b}\approx\sqrt{2S}\left(1-\frac{\hc{b}\op{b}}{4S}\right)\op{b},\\
    \label{eq:HP_Sminus}
    \op{S}_{-}&=\op{S}_{x}-i\op{S}_{y}=\hc{b}\left(\sqrt{2S-\hc{b}\op{b}}\right)\approx\sqrt{2S}\hc{b}\left(1-\frac{\hc{b}\op{b}}{4S}\right),
\end{align}
which gives
\begin{align}
    \label{eq:HP_Sx}
    \op{S}_{x}&=\sqrt{\frac{S}{2}}\left[\op{b}-\frac{\hc{b}\op{b}\op{b}}{4S}+\hc{b}-\frac{\hc{b}\hc{b}\op{b}}{4S}\right],\\
    \label{eq:HP_Sy}
    \op{S}_{y}&=-i\sqrt{\frac{S}{2}}\left[\op{b}-\frac{\hc{b}\op{b}\op{b}}{4S}-\hc{b}+\frac{\hc{b}\hc{b}\op{b}}{4S}\right],\\
    \label{eq:HP_Sz}
    \op{S}_{z}&=S-\hc{b}\op{b}.
\end{align}
where $\hat{b}^\dagger (\hat{b})$ are the creation (annihilation) operators of the magnon mode. The Hamiltonian is thus transformed into
\begin{align}
    \label{eq:Hamiltonian_oop_HP}
    \op{H}_\mathrm{OOP} = \gamma\left( B_\mathrm{ext} + \frac{2 K_\mathrm{u}}{M_\mathrm{s}}\right)\hc{b}\op{b}-\frac{\gamma^2 K_\mathrm{u}}{M_\mathrm{s}^2V_\mathrm{m}} \hc{b}\op{b}\hc{b}\op{b}.
\end{align}
We set $\op{m} = \op{b}/\sqrt{\hbar}$, $\hc{m}=\hc{b}/\sqrt{\hbar}$ for the magnon (dimensionless) operators. The previous condition for linearization $\hc{b}\op{b}\ll{}2S$ becomes $\hc{m}\op{m}\ll{}N$, and we write
\begin{align}
    \label{eq:Hamiltonian_oop_harm}
    \op{H}_{\rm{OOP}}^{(\rm{h})} &= \hbar\gamma\left( B_\mathrm{ext} + \frac{2 K_\mathrm{u}}{M_\mathrm{s}}\right)\left(\hc{m}\op{m}+\frac{1}{2}\right)\\&=\hbar \omega_\mathrm{m}\left(\hc{m}\op{m}+\frac{1}{2}\right)
\end{align}
as the harmonic part of the Hamiltonian, where $\omega_\mathrm{m}=\gamma( B_\mathrm{ext} -B_\mathrm{eff})$ is the OOP Kittel relation, with $B_{\rm{eff}}=-2K_{\rm{u}}/M_{\rm{s}}$. This definition of $B_{\rm{eff}}$ is consistent with the definition of the effective magnetization in ferromagnetic resonance and hence it is different in sign from $\vect{H}_{\rm{a}}$.
The anharmonic part of the Hamiltonian is written as
\begin{align}
    \label{eq:Hamiltonian_oop_anharm}
    \op{H}_{\rm{OOP}}^{(\rm{nh})}=-\frac{\gamma^2K_{\rm{u}}\hbar^2}{M_{\rm{s}}^2V_{\rm{m}}} \hc{m}\op{m}\hc{m}\op{m}=\hbar\mathcal{K}_\mathrm{OOP} \hc{m}\op{m}\hc{m}\op{m}
\end{align}
where we introduced the OOP Kerr coefficient 
\begin{equation}
    \label{eq:oop_Kerr}
    \mathcal{K}_\mathrm{OOP}=-\frac{\gamma^2K_{\rm{u}}\hbar}{M_{\rm{s}}^2V_{\rm{m}}}.
\end{equation}
If we consider that the effective magnetic anisotropy only arises from the shape anisotropy, i.e., $K_\mathrm{u}=-\mu_0M_\mathrm{s}^2/2$, we obtain
\begin{equation}
    \label{eq:oop_Kerr_OOP_simplified}
    \mathcal{K}_\mathrm{OOP}=\frac{\mu_0\gamma^2\hbar}{2 V_{\rm{m}}}.
\end{equation}
The OOP Kerr coefficient is then material- and temperature-independent, as it scales solely with the volume of the magnetic system, but does not depend on $M_{\rm{s}}$.

\subsubsection{In-plane field}

Here, we consider an IP external field (defining the new direction $z$), and a uniaxial anisotropy $K_{ \rm{u}}=K_{\rm{a}}$ with $\vect{e}_{\rm{a}}$ now along $x$. Because $\vect{H}_{\rm{ext}}$ and $\vect{e}_{\rm{a}}$ are not aligned, the magnetic Hamiltonian $\op{H}_\mathrm{IP}$ is written as
\begin{equation}\label{eq:Hamiltonian_ip}
    \op{H}_\mathrm{IP} = -\gamma{}B_\mathrm{ext}\op{S}_z-\frac{\gamma^2K_{\rm{u}}}{M_{\rm{s}}^2V_{\rm{m}}}\op{S}_x^2.
\end{equation}
We consider $\vect{H}_{\rm{ext}}$ to be sufficiently large such that the static $\vect{M}$ is in the plane along $z$ and neglect further anisotropy terms (e.g., the cubic $K_1$) that could tilt $\vect{M}$ away from $z$.
We use that $\op{S}_x^2+\op{S}_y^2+\op{S}_z^2$ expanded as Eqs.~\eqref{eq:HP_Sx2}--\eqref{eq:HP_Sz2} is a constant to obtain within the conditions of the Holstein-Primakoff approximation
\begin{equation}\label{eq:Hamiltonian_ip2}
    \op{H}_\mathrm{IP} = -\gamma{}B_\mathrm{ext}\op{S}_z+\frac{\gamma^2K_{\rm{u}}}{2M_{\rm{s}}^2V_{\rm{m}}}\op{S}_z^2-\frac{\gamma^2K_{\rm{u}}}{2M_{\rm{s}}^2V_{\rm{m}}}(\op{S}_x^2-\op{S}_y^2).
\end{equation}
The Hamiltonian is then transformed into
\begin{multline}\label{eq:Hamiltonian_ip_HP}
    \op{H}_\mathrm{IP} = \gamma{}B_\mathrm{ext}\hc{b}\op{b}
    -\frac{\gamma{}K_{\rm{u}}}{M_{\rm{s}}}\hc{b}\op{b}    -\frac{\gamma{}K_{\rm{u}}}{2M_{\rm{s}}}\left[    \op{b}\op{b}+\hc{b}\hc{b}\right]\\
    +\frac{\gamma^2K_{\rm{u}}}{2M_{\rm{s}}^2V_{\rm{m}}}\left[\hc{b}\op{b}\hc{b}\op{b}+\frac{\hc{b}\op{b}\op{b}\op{b}+\op{b}\hc{b}\op{b}\op{b}+\hc{b}\hc{b}\op{b}\hc{b}+\hc{b}\hc{b}\hc{b}\op{b}}{4}
    \right].
\end{multline}
To factorize the second-order terms into $\hc{b}\op{b}$ only, we introduce
\begin{align}
    \label{eq:Bog_A}
    2\mathcal{A}&=\gamma{}B_\mathrm{ext}-\frac{\gamma{}K_{\rm{u}}}{M_{\rm{s}}},\\
    \label{eq:Bog_B}
    \mathcal{B}&=-\frac{\gamma{}K_{\rm{u}}}{2M_{\rm{s}}},
\end{align}
and proceed with the Bogoliubov transformation of the annihilation and creation operators. We introduce $\op{m},\hc{m}$ such that
\begin{equation}
    \label{eq:Bog_b_bdag}
    \op{b}/\sqrt{\hbar}=u\op{m}-v\hc{m},\ \hc{b}/\sqrt{\hbar}=u\hc{m}-v\op{m},
\end{equation}
and
\begin{equation}
    \label{eq:Bog_m_mdag}
    \sqrt{\hbar}\op{m}=\frac{u\op{b}+v\hc{b}}{u^2-v^2},\ 
    \sqrt{\hbar}\hc{m}=\frac{u\hc{b}+v\op{b}}{u^2-v^2}.
\end{equation}
Preserving the canonical $[\op{b},\hc{b}]=\hbar$ into $[\op{m},\hc{m}]=1$ requires $u^2-v^2=1$. To eliminate terms in $\op{m}\op{m}$ and $\hc{m}\hc{m}$ in the transformed Hamiltonian, it is required that $\mathcal{B}(u^2+v^2)=2uv\mathcal{A}$, that is, $(u^2+v^2)=2uv\mathcal{R}$ with $\mathcal{R}=\mathcal{A}/\mathcal{B}=1-B_\mathrm{ext}M_{\rm{s}}/K_{\rm{u}}$. Since $K_{\rm{u}}<0$ in the case of the magnetostatic anisotropy of a thin film, $\mathcal{R}>1$, and a solution is
\begin{equation}
    \label{eq:Bog_u_v_sols}
    u=\frac{1}{\sqrt{2}}\sqrt{\frac{\mathcal{R}}{\sqrt{\mathcal{R}^2-1}}+1},\
    v=\frac{1}{\sqrt{2}}\sqrt{\frac{\mathcal{R}}{\sqrt{\mathcal{R}^2-1}}-1}.
\end{equation}
Hence, the harmonic part of $\op{H}_{\rm{IP}}$ is 
\begin{equation}
    \label{eq:Hamiltonian_ip_Bog_harm}
    \begin{split}
    \op{H}_{\rm{IP}}^{(\rm{h})}&=\hbar\left[\mathcal{A}(u^2+v^2)-2uv\mathcal{B}\right](\hc{m}\op{m}+\op{m}\hc{m})\\
    &=\hbar\gamma\sqrt{B_\mathrm{ext}\left(B_\mathrm{ext}-\frac{2K_{\rm{u}}}{M_{\rm{s}}}\right)}(\hc{m}\op{m}+\frac{1}{2}),
    \end{split}
\end{equation}
with a constant term that is ignored. This form provides the resonance angular frequency of the magnon system, which is $\omega_{\rm{m}}=\gamma[B_\mathrm{ext}(B_\mathrm{ext}+B_{\rm{eff}})]^{1/2}$. It corresponds to the IP Kittel formula, with $B_{\rm{eff}}=-2K_{\rm{u}}/M_{\rm{s}}$ as above.

The other relevant part of the Hamiltonian is the anharmonic part
\begin{equation}
    \label{eq:Hamiltonian_ip_HP_anharm}
    \begin{split}
    &\op{H}_{\rm{IP}}^{(\rm{nh})}\\&=\mathcal{C}\left[\hc{b}\op{b}\hc{b}\op{b}+\frac{\hc{b}\op{b}\op{b}\op{b}+\op{b}\hc{b}\op{b}\op{b}+\hc{b}\hc{b}\op{b}\hc{b}+\hc{b}\hc{b}\hc{b}\op{b}}{4}
    \right],
    \end{split}
\end{equation}
with $\mathcal{C}=\gamma^2K_{\rm{u}}/(2M_{\rm{s}}^2V_{\rm{m}})$. We note $\op{S}_1=\op{m}\hc{m}+\hc{m}\op{m}$ and $\op{S}_2=\hc{m}\hc{m}+\op{m}\op{m}$. Then,
\begin{equation}
    \label{eq:Hamiltonian_ip_Bog_anharm}
    \begin{split}
    \op{H}_{\rm{IP}}^{(\rm{nh})}
    &= \hbar^2\mathcal{C}\left\{ \vphantom{\frac{1}{1}} \right. 
    u^4\left[\hc{m}\op{m}\hc{m}\op{m}
        + \op{S}_1\op{m}\op{m}/4
        + \hc{m}\hc{m}\op{S}_1/4\right]\\
    &+ v^4\left[\op{m}\hc{m}\op{m}\hc{m}
        + \op{S}_1\hc{m}\hc{m}/4
        + \op{m}\op{m}\op{S}_1/4\right]\\
    &+ u^2v^2\left[\hc{m}\op{m}\op{m}\hc{m}
        + \op{m}\hc{m}\hc{m}\op{m}
        + \op{S}_2^2\right.\\
    &\left. +\, 3(\op{S}_1\op{S}_2+\op{S}_2\op{S}_1)/4\right]
    - u^3v\left[\op{S}_1^2/2
        + \op{S}_2\op{m}\op{m}/2\right.\\
    &\left.
        + \hc{m}\hc{m}\op{S}_2/2
        + \hc{m}\op{m}\op{S}_2+\op{S}_2\hc{m}\op{m}\right]
    - uv^3\left[\op{S}_1^2/2\right.\\
    &\left.
        + \op{S}_2\hc{m}\hc{m}/2
        + \op{m}\op{m}\op{S}_2/2
        + \op{m}\hc{m}\op{S}_2
        + \op{S}_2\op{m}\hc{m}\right]
    \left.\vphantom{\frac{1}{1}} \right\}.
    \end{split}
\end{equation}
Employing the rotating wave approximation, fast oscillating terms with unequal number of $\op{m}$ and $\hc{m}$ operators are omitted. As only the terms of the correct order with two of each $\op{m},\hc{m}$ operators are kept, Eq.~\eqref{eq:Hamiltonian_ip_Bog_anharm} simplifies into
\begin{equation}
    \label{eq:Hamiltonian_ip_Bog_anharmfilt}
    \begin{split}
    \op{H}_{\rm{IP}}^{(\rm{nh})}=\hbar^2\mathcal{C}&\left[
    \left(u^2v^2-u^3v\right)\hc{m}\hc{m}\op{m}\op{m}\right.\\
    &+\left(u^4-u^3v/2-uv^3/2\right)\hc{m}\op{m}\hc{m}\op{m}\\
    &+\left(u^2v^2-u^3v/2-uv^3/2\right)\hc{m}\op{m}\op{m}\hc{m}\\
    &+\left(u^2v^2-u^3v/2-uv^3/2\right)\op{m}\hc{m}\hc{m}\op{m}\\
    &+\left(v^4-u^3v/2-uv^3/2\right)\op{m}\hc{m}\op{m}\hc{m}\\
    &+\left.\left(u^2v^2-uv^3\right)\op{m}\op{m}\hc{m}\hc{m}
    \right],
    \end{split}
\end{equation}
which can be reordered as 
\begin{equation}
    \label{eq:Hamiltonian_ip_Bog_short}
    \begin{split}
    \op{H}_{\rm{IP}}^{(\rm{nh})}=\hbar^2\mathcal{C}&\left[
    (u-v)^2(u^2-uv+v^2)\hc{m}\op{m}\hc{m}\op{m}\right.\\
    &+\left.(u-v)^2(2v-u)v\hc{m}\op{m}
    \right],
    \end{split}
\end{equation}
corresponding to
\begin{equation}
    \label{eq:Hamiltonian_ip_Bog_vals}
    \begin{split}
    \op{H}_{\rm{IP}}^{(\rm{nh})}&=\hbar^2\frac{\gamma^2K_{\rm{u}}}{2M_{\rm{s}}^2V_{\rm{m}}}\Bigg( \frac{B_{\rm{eff}}/4+B_\mathrm{ext}}{B_{\rm{eff}}+B_\mathrm{ext}}\hc{m}\op{m}\hc{m}\op{m}\\
    &\hphantom{=}+\frac{B_{\rm{eff}}/4+B_\mathrm{ext}-\sqrt{B_\mathrm{ext}(B_{\rm{eff}}+B_\mathrm{ext})}}{B_{\rm{eff}}+B_\mathrm{ext}}\hc{m}\op{m}\Bigg).   \end{split}
\end{equation}
In \cref{eq:Hamiltonian_ip_Bog_vals}, the second term in $\hc{m}\op{m}$ can be neglected, since it is in the order of \si{nHz}, much smaller than $\omega_{\rm{m}}=\gamma[B_\mathrm{ext}(B_\mathrm{ext}+B_{\rm{eff}})]^{1/2}$ in \cref{eq:Hamiltonian_ip_Bog_harm}. Although this second term in $\hc{m}\op{m}$ can always be neglected, the term in $\hc{m}\op{m}\hc{m}\op{m}$ becomes significant with increasing magnon population as $\hc{m}\op{m}\hc{m}\op{m}\gg\hc{m}\op{m}$, which can no longer be neglected for $\hc{m}\op{m}\gg1$ and gives rise to the nonlinearity. Note, however, that the condition $\hc{m}\op{m}\ll{}N$ is still required by the Holstein-Primakoff transformation, which applies to most cases of practical interest, as we illustrate in Sec.~\ref{sec:results}. In the situation where the external field and anisotropy axis are orthogonal to each other, we therefore obtain the IP Kerr coefficient 
\begin{equation}
    \label{eq:ip_Kerr}
    \mathcal{K}_\mathrm{IP}=\xi\frac{\gamma^2K_{\rm{u}}\hbar}{2M_{\rm{s}}^2V_{\rm{m}}}
\end{equation}
where 
\begin{equation}
\label{eq:xi}
    \xi=\frac{B_{\rm{eff}}/4+B_\mathrm{ext}}{B_{\rm{eff}}+B_\mathrm{ext}}
\end{equation} 
is a geometric factor originating from the ellipticity of the magnetic precession. Using the inverse Kittel relation 
 \begin{equation}
    \label{eq:inverse_Kittel_IP}
    B_\mathrm{ext} = -B_\mathrm{eff}/2+\sqrt{(B_\mathrm{eff}/2)^2+(\omega_\mathrm{m}/\gamma)^2},
\end{equation}
we can obtain the dependency of $\xi$ on $\omega_\mathrm{m}$
\begin{equation}
    \label{eq:xi_vs_omega_m}
    \xi=\frac{\sqrt{(B_\mathrm{eff}/2)^2+(\omega_\mathrm{m}/\gamma)^2}-B_\mathrm{eff}/4}{\sqrt{(B_\mathrm{eff}/2)^2+(\omega_\mathrm{m}/\gamma)^2}+B_\mathrm{eff}/2}
\end{equation}
In the simplest case where $K_\mathrm{u}=-\mu_0M_\mathrm{s}^2/2$, we obtain
\begin{equation}
\label{eq:Kerr_coeff_IP_simplified}
    \mathcal{K}_\mathrm{IP} = -\xi \frac{\mu_0 \gamma^2 \hbar}{4 V_\mathrm{m}} = -\frac{\xi}{2}\mathcal{K}_\mathrm{OOP}.
\end{equation}
In contrast to the OOP case, the IP Kerr coefficient then depends on the magnetic material and temperature through the ellipticity term $\xi$, while the scaling with the inverse volume remains similar.

\subsubsection{Intermediate field angles}

Here, we consider an external field $\vect{H}_{\rm{ext}}=H_{\rm{ext}}\,(\sin{\theta}\,\vect{e}_{\rm{x}}+\cos{\theta}\,\vect{e}_{\rm{z}})$ oriented at an arbitrary angle $\theta$ between the OOP and IP directions, and a uniaxial anisotropy $K_{ \rm{u}}=K_{\rm{a}}$ with $\vect{e}_{\rm{a}}$ along $z$. Because $\vect{H}_{\rm{ext}}$ is not along the direction of an energy maximum or minimum for the anisotropy, the magnetization and external magnetic field are not aligned. The classical energies of the magnetic system are 
\begin{equation}\label{eq:Hamiltonian_start_theta}
    H_\mathrm{mag} = -V_{\rm{m}}\left[\mu_0H_{\rm{ext}}M_{\rm{s}}\cos(\theta_{\rm{M}}-\theta) + K_{\rm{u}}\cos^2(\theta_{\rm{M}})\right],
\end{equation}
with $\theta_{\rm{M}}$ the angle between $\vect{m}$ and $\vect{e}_{\rm{z}}$. The equilibrium angle $\theta_{\rm{M}}$ is found by the zero of the derivative of the energy $\mathrm{d}H_\mathrm{mag}/\mathrm{d}\theta_{\rm{M}}$, which provides $\mu_0H_{\rm{ext}}M_{\rm{s}}\sin(\theta_{\rm{M}}-\theta)=-2K_{\rm{u}}\cos\theta_{\rm{M}}\sin\theta_{\rm{M}}$, or
\begin{equation}\label{eq:Hamiltonian_equilibrium}
\gamma{}B_{\rm{ext}}\sin(\theta_{\rm{M}}-\theta)=-4\mathcal{C}S\cos\theta_{\rm{M}}\sin\theta_{\rm{M}}.
\end{equation}
The magnetic Hamiltonian $\op{H}_\mathrm{\theta}$ is written as
\begin{equation}\label{eq:Hamiltonian_theta_start}
    \op{H}_\mathrm{\theta} = -\gamma{}B_\mathrm{ext}\sin{\theta}\op{S}_x-\gamma{}B_\mathrm{ext}\cos{\theta}\op{S}_z-\frac{\gamma^2K_{\rm{u}}}{M_{\rm{s}}^2V_{\rm{m}}}\op{S}_z^2,
\end{equation}
which we need to transform using
\begin{align}\label{eq:Sm}
    \op{S}_m&=\sin{\theta_{\rm{M}}}\op{S}_x+\cos{\theta_{\rm{M}}}\op{S}_z,\\
    \label{eq:Sn}
    \op{S}_n&=\cos{\theta_{\rm{M}}}\op{S}_x-\sin{\theta_{\rm{M}}}\op{S}_z,
\end{align}
in order to obtain
\begin{equation}\label{eq:Hamiltonian_theta_SmSn}
    \begin{split}
    \op{H}_\mathrm{\theta} =& -\gamma{}B_\mathrm{ext}\cos(\theta_{\rm{M}}-\theta)\op{S}_m+\gamma{}B_\mathrm{ext}\sin(\theta_{\rm{M}}-\theta)\op{S}_n\\
    &+\frac{\gamma^2K_{\rm{u}}}{M_{\rm{s}}^2V_{\rm{m}}}\left[\cos\theta_{\rm{M}}\sin\theta_{\rm{M}}(\op{S}_m\op{S}_n+\op{S}_n\op{S}_m)\right]\\
    &-\frac{\gamma^2K_{\rm{u}}}{M_{\rm{s}}^2V_{\rm{m}}}\left[\cos^2(\theta_{\rm{M}})\op{S}_m^2+\sin^2(\theta_{\rm{M}})\op{S}_n^2\right].
    \end{split}
\end{equation}
We perform the Holstein-Primakoff transformation of the total spin operator in this new frame of reference. Injecting the equilibrium condition of Eq.~\eqref{eq:Hamiltonian_equilibrium}, the terms in $\op{b},\hc{b}$ vanish and Eq.~\eqref{eq:Hamiltonian_theta_SmSn} becomes (see \Cref{sec:rotated_HP})
\begin{equation}
    \label{eq:Hamiltonian_theta_simpHP}
    \begin{split}
    \op{H}_\mathrm{\theta} =& \left[\gamma{}B_\mathrm{ext}\cos(\theta_{\rm{M}}-\theta)+4\mathcal{C}S(\cos^2\theta_{\rm{M}}-\sin^2\theta_{\rm{M}}/2)\right]\hc{b}\op{b}\\
    &-\mathcal{C}S\sin^2\theta_{\rm{M}}\left[\op{b}\op{b}+\hc{b}\hc{b}\right]\\
        &-2\mathcal{C}\cos\theta_{\rm{M}}\sin\theta_{\rm{M}}\sqrt{\frac{S}{2}}(\hc{b}\op{b}\op{b}+\hc{b}\op{b}\hc{b}+\op{b}\hc{b}\op{b}+\hc{b}\hc{b}\op{b})\\
        &+\mathcal{C}\sin^2\theta_{\rm{M}}\left[\frac{\hc{b}\op{b}\op{b}\op{b}+\op{b}\hc{b}\op{b}\op{b}+\hc{b}\hc{b}\op{b}\hc{b}+\hc{b}\hc{b}\hc{b}\op{b}}{4}\right]\\
    &-2\mathcal{C}(\cos^2\theta_{\rm{M}}-\sin^2\theta_{\rm{M}}/2)(\hc{b}\op{b}\hc{b}\op{b}).
    \end{split}
\end{equation}
Similar to the IP case, the second-order terms $\op{b}\op{b},\hc{b}\hc{b}$ can be eliminated with the Bogoliubov transformation. We introduce
\begin{align}
    \label{eq:Bog_A_theta}
    2\mathcal{A}&=\gamma{}\left[B_\mathrm{ext}\cos(\theta_{\rm{M}}-\theta)+\frac{K_{\rm{u}}}{M_{\rm{s}}}\left(2\cos^2\theta_{\rm{M}}-\sin^2\theta_{\rm{M}}\right)\right],\\
    \label{eq:Bog_B_theta}
    \mathcal{B}&=-\frac{\gamma{}K_{\rm{u}}}{2M_{\rm{s}}}\sin^2\theta_{\rm{M}},
\end{align}
and for the intermediate angles, we define
\begin{equation}
\begin{split}
    \label{eq:R_theta}
    \mathcal{R}=\mathcal{A}/\mathcal{B}&=1-\frac{2\cos^2\theta_{\rm{M}}}{\sin^2\theta_{\rm{M}}}-\frac{B_\mathrm{ext}M_{\rm{s}}\cos(\theta_{\rm{M}}-\theta)}{K_{\rm{u}}\sin^2\theta_{\rm{M}}}\\
    &=1+\frac{2\cos\theta_{\rm{M}}}{\sin\theta_{\rm{M}}}\left[\frac{\cos(\theta_{\rm{M}}-\theta)}{\sin(\theta_{\rm{M}}-\theta)}-\frac{\cos\theta_{\rm{M}}}{\sin\theta_{\rm{M}}}\right]
\end{split}
\end{equation}
with $\mathcal{R}>1$. We use again the dimensionless $\op{m},\hc{m}$ operators with $u,v$ coefficients defined using $\mathcal{R}$ from Eq.~\eqref{eq:R_theta} as in Eqs.~\eqref{eq:Bog_b_bdag},\eqref{eq:Bog_m_mdag}. The harmonic part of the Hamiltonian is obtained as
\begin{equation}
    \label{eq:Hamiltonian_theta_Bog_harm}
    \begin{split}
    \op{H}_{\theta}^{(\rm{h})}&=\hbar\left[(2\mathcal{A}+2\mathcal{B})(2\mathcal{A}-2\mathcal{B})\right]^{1/2}(\hc{m}\op{m}+\frac{1}{2})\\
    &=\hbar\gamma\left[\left(B_\mathrm{ext}\cos(\theta_{\rm{M}}-\theta)+\frac{2K_{\rm{u}}}{M_{\rm{s}}}\cos^2\theta_{\rm{M}}\right)\right.\\
    &\hphantom{=}\left.\left(B_\mathrm{ext}\cos(\theta_{\rm{M}}-\theta)+\frac{2K_{\rm{u}}}{M_{\rm{s}}}\cos(2\theta_{\rm{M}})\right)\right]^{1/2}\\&\hphantom{=}(\hc{m}\op{m}+\frac{1}{2}),
    \end{split}
\end{equation}
which corresponds to $\omega_{\rm{m}}$ for an arbitrary angle of the external field.

The other relevant part of the Hamiltonian is its anharmonic part
\begin{equation}
    \label{eq:Hamiltonian_theta_HP_anharm}
    \begin{split}
    \op{H}_{\theta}^{(\rm{nh})}&=\mathcal{C}\sin^2\theta_{\rm{M}}\frac{\hc{b}\op{b}\op{b}\op{b}+\op{b}\hc{b}\op{b}\op{b}+\hc{b}\hc{b}\op{b}\hc{b}+\hc{b}\hc{b}\hc{b}\op{b}}{4}\\
    &\hphantom{=}-2\mathcal{C}\cos\theta_{\rm{M}}\sin\theta_{\rm{M}}\sqrt{\frac{S}{2}}\left(\hc{b}\op{b}\op{b}+\hc{b}\op{b}\hc{b}+\op{b}\hc{b}\op{b}+\hc{b}\hc{b}\op{b}\right)\\
    &\hphantom{=}-2\mathcal{C}(\cos^2\theta_{\rm{M}}-\sin^2\theta_{\rm{M}}/2)(\hc{b}\op{b}\hc{b}\op{b}),
    \end{split}
\end{equation}
in which the terms in $\hc{b}\op{b}\op{b},\hc{b}\op{b}\hc{b},\op{b}\hc{b}\op{b},\hc{b}\hc{b}\op{b}$ correspond to the shift of equilibrium position for $\vect{m}$ with excitation power, due to the non-linear anisotropy evolution with excitation angle. Ignoring this correction to $\theta_\mathrm{M}$ (up to $\approx0.5{^\circ}$ for typical parameters introduced below), we can transform $\op{b},\hc{b}$ into $\op{m},\hc{m}$ and only keep terms of the correct order in the rotating wave approximation to obtain
\begin{equation}
    \label{eq:Hamiltonian_theta_Bog_short}
    \begin{split}
    \op{H}_{\theta}^{(\rm{nh})}=\hbar^2\mathcal{C}&\left[
    (u-v)^2(u^2-uv+v^2)\sin^2\theta_{\rm{M}}\right.\\
    &-\left.2(1+6u^2v^2)\cos^2\theta_{\rm{M}}\right]\hc{m}\op{m}\hc{m}\op{m},
    \end{split}
\end{equation}
and some terms in $\hc{m}\op{m}$ with a prefactor in the range of \si{nHz} that can be neglected, as was the case for Eq.~\eqref{eq:Hamiltonian_ip_Bog_vals}. Note that the main term of Eq.~\eqref{eq:Hamiltonian_theta_Bog_short} cannot be neglected since $\hc{m}\op{m}\hc{m}\op{m}\gg\hc{m}\op{m}$. This corresponds to
\begin{equation}
    \label{eq:Hamiltonian_theta_Bog_vals}
    \begin{split}
    \op{H}_{\theta}^{(\rm{nh})}&=\hbar^2\frac{\gamma^2K_{\rm{u}}}{2M_{\rm{s}}^2V_{\rm{m}}}\left[\left(1-\frac{3}{2(\mathcal{R}+1)}\right)\sin^2\theta_{\rm{M}}\right.\\
    &\hphantom{=}-\left.\left(2+\frac{3}{\mathcal{R}^2-1}\right)\cos^2\theta_{\rm{M}}\right]\hc{m}\op{m}\hc{m}\op{m},
    \end{split}
\end{equation}
which provides the Kerr coefficient
\begin{equation}
    \label{eq:theta_Kerr}
    \begin{split}
    \mathcal{K}_\theta&=\frac{\gamma^2K_{\rm{u}}\hbar}{2M_{\rm{s}}^2V_{\rm{m}}}\left(\xi'\sin^2\theta_{\rm{M}}-2\xi''\cos^2\theta_{\rm{M}}\right)\\
    &=(\xi''\cos^2\theta_{\rm{M}}-\xi'\sin^2\theta_{\rm{M}}/2)\mathcal{K}_{\rm{OOP}},
    \end{split}
\end{equation}
where 
\begin{equation}
\label{eq:xip_xipp}
    \xi'=1-\frac{3}{2(\mathcal{R}+1)},\,\xi''=1+\frac{3}{2(\mathcal{R}+1)(\mathcal{R}-1)}
\end{equation} 
are geometric factors originating from the ellipticity of the magnetic precession. In the limit of $\theta_\mathrm{M}=\SI{0}{\degree}$, $\mathcal{R}\rightarrow\infty$ (with $u\rightarrow1,v\rightarrow0$) and $\mathcal{K}_\theta=\mathcal{K}_{\rm{OOP}}$. In the limit of $\theta_\mathrm{M}=\SI{90}{\degree}$, $\xi'=\xi$ and $\mathcal{K}_\theta=\mathcal{K}_{\rm{IP}}$.

A particular case is the angle for which the cancellation of the Kerr coefficient occurs. In the limit of large external fields, where $\theta_\mathrm{M}$ approaches $\theta$, $\mathcal{R}\rightarrow\infty$ with an almost circular magnetic precession and $\xi'\rightarrow1,\,\xi''\rightarrow1$, so that the MKE cancels for $\theta_{\rm{M}}=\arctan{\sqrt{2}}\approx\SI{54.74}{\degree}$, the magic angle. At lower fields, the ellipticity of the magnetic precession with $\xi'<1,\,\xi''>1$ shifts this cancellation angle towards larger values. A complete angle-dependence of $\mathcal{R}$, $\xi'$, $\xi''$, and $\mathcal{K}_\theta$, computed for YIG, is presented in \cref{sec:kerr_angle_simulation}.

In \cref{eq:Hamiltonian_theta_simpHP}, we used the equilibrium condition for $\theta_\mathrm{M}$ given by \cref{eq:Hamiltonian_equilibrium}. For a large magnon population, because the angular dependence of $H_\mathrm{mag}$ is not symmetric along the precession orbit, the equilibrium magnetization angle dynamically shifts to minimize the total magnetic energy. To quantitatively account for this effect, we now start from \cref{eq:Hamiltonian_theta_SmSn} without employing the equilibrium condition \cref{eq:Hamiltonian_equilibrium}, and only deal with the terms in first and third order of the magnon operators. Using the Holstein-Primakoff transformation [\cref{eq:HP_Sn,eq:HP_Syp,eq:HP_Sm}] yields
\begin{equation}
    \label{eq:Hamiltonian_theta_HP_third}
    \begin{split}
    \op{H}_{\theta}^{(\rm{dir})}=&\hphantom{+}\gamma B_\mathrm{ext}\sin{(\theta_\mathrm{M}'-\theta)}\sqrt{\frac{S}{2}}\left(\op{b}+\hc{b}-\frac{\hc{b}\op{b}\op{b}+\hc{b}\hc{b}\op{b}}{4S}\right)\\
    &+\sqrt{2S}\mathcal{C}\cos{\theta_\mathrm{M}'}\sin{\theta_\mathrm{M}'}\left[\vphantom{\frac{3}{2}}2S(\op{b}+\hc{b})\right.\\
    &\left.-\frac{3}{2}\hc{b}\op{b}\op{b}-\frac{3}{2}\hc{b}\hc{b}\op{b}-\hc{b}\op{b}\hc{b}-\op{b}\hc{b}\op{b}\right],
    \end{split}
\end{equation}
where $\theta_\mathrm{M}'$ is the new equilibrium magnetization angle under a large magnon population. After transformation into $\hc{m},\op{m}$ and removing fast-oscillating terms $\hc{m}\hc{m}\hc{m},\op{m}\op{m}\op{m}$ in the rotating wave approximation, this becomes
\begin{equation}
    \label{eq:Hamiltonian_theta_Bog_third}
    \begin{split}
    \op{H}_{\theta}^{(\rm{dir})}=\left[\gamma B_\mathrm{ext}\sin{(\theta_\mathrm{M}'-\theta)}\sqrt{\frac{S}{2}}+(2S)^{3/2}\mathcal{C}\cos{\theta_\mathrm{M}'}\sin{\theta_\mathrm{M}'}\right]\\
    \sqrt{\hbar}(u-v)(\op{m}+\hc{m})\\
    -\left[\gamma B_\mathrm{ext}\sin{(\theta_\mathrm{M}'-\theta)}\frac{1}{4\sqrt{2S}}+3\sqrt{\frac{S}{2}}\mathcal{C}\cos{\theta_\mathrm{M}'}\sin{\theta_\mathrm{M}'}\right]\\
    \hbar^{3/2}\left[u^3(\hc{m}\op{m}\op{m}+\hc{m}\hc{m}\op{m})-v^3(\op{m}\hc{m}\hc{m}+\op{m}\op{m}\hc{m})\right.\\
    \left.+uv^2(\op{m}\op{S}_1+\op{S}_1\hc{m})-u^2v(\hc{m}\op{S}_1+\op{S}_1\op{m})\right]\\
    -\sqrt{2S}\mathcal{C}\cos{\theta_\mathrm{M}'}\sin{\theta_\mathrm{M}'}\hbar^{3/2}\left[(u^3-v^3)(\hc{m}\op{m}\hc{m}+\op{m}\hc{m}\op{m})\right.\\
    \left.+uv(v-u)(\hc{m}\op{m}\op{m}+\op{m}\op{m}\hc{m}+\op{m}\hc{m}\hc{m}+\hc{m}\hc{m}\op{m})\right].
    \end{split}
\end{equation}
For large magnon numbers, we will show in \cref{sec:eom} that \cref{eq:Hamiltonian_theta_Bog_third} leads to a condition on $\theta_\mathrm{M}'$ ensuring that the Hamiltonian $\op{H}_{\theta}^{(\rm{dir})}$ vanishes for a given mode population $\hc{m}\op{m}$, which determines the shifted equilibrium and can be obtained numerically. For the simulations presented in \cref{sec:results}, $\langle \hat{m}^\dagger \hat{m} \rangle/N\approx0.01$ and the correction to $\theta_\mathrm{M}$ is in the order of \SI{0.5}{\degree}, and could be neglected. However, this gradual reorientation of $\theta_\mathrm{M}'$ induces a frequency shift that grows linearly with the magnon number. Therefore, it constitutes an anharmonicity arising indirectly from the third order terms in $\hc{m},\op{m}$ and that offsets the self-Kerr anharmonicity directly coming from the fourth order term $\mathcal{K}_{\theta}\,\hc{m}\op{m}\hc{m}\op{m}$. This correction to the total anharmonicity occurs as soon as the external field is applied neither in the plane nor out of the plane. Such a correction to the Kerr coefficient exhibits a maximum in absolute value at a specific angle $\theta$ between 0 and \SI{90}{\degree}, and is more pronounced for low values of the external field.

\subsection{Equations of motion for a hybrid cavity--magnonics system}\label{sec:eom}

Cavity magnonics has become a widely used platform for studying magnonic phenomena in magnetic materials \cite{ZARERAMESHTI20221, PhysRevLett.111.127003, PhysRevLett.113.083603, PhysRevLett.113.156401, PhysRevLett.123.107701, PhysRevLett.123.107702, Guo2023}. The cavity resonance mode serves as a sensitive probe of magnetization dynamics. Moreover, the strong microwave field confinement in the cavity can efficiently drive the magnonic system to nonlinear regimes with moderate drive powers, making the cavity an important driving and probing tool for hybridized systems. Here we detail how to obtain the EOM for the Hamiltonian describing the hybrid cavity--magnon system with the nonlinear magnon Kerr term in the classical limit. We consider a cavity that is entirely linear (perfectly harmonic oscillator), loaded with the sample, and connected via two microwave ports. By using input-output theory, we obtain magnon and photon populations which describe the dynamics of the system \cite{RevModPhys.86.1391}. 

\begin{figure}[t]
\includegraphics[width=\linewidth]{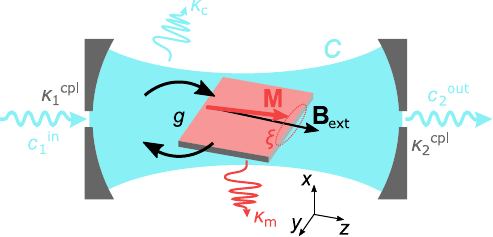}%
\caption{Thin ferromagnetic film magnetized IP in a microwave cavity. The magnetization $\mathbf{M}$ of the film precesses elliptically around the external field $\mathbf{B}_\mathrm{ext}$, with the ellipticity parameter $\xi$, and frequency given by the photonic excitation. The ferromagnetic resonance couples to the cavity mode with the coupling constant $g$. The cavity has two ports, with incoming ($c_1^\mathrm{in}$) and outgoing ($c_2^\mathrm{out}$) photon amplitudes, and $\kappa_1^\mathrm{cpl}$ and $\kappa_2^\mathrm{cpl}$ are the coupling coefficients for respective port. The cavity has an internal dissipation rate $\kappa_\mathrm{c}$, and photon amplitude $C$. The magnon dissipation rate is given by $\kappa_\mathrm{m}$.}
\label{fig:fig_setup}
\end{figure}

The Hamiltonian of the hybrid system is given by
\begin{equation}\label{eq:Hamiltonian_full}
    \hat{H}/\hbar = \omega_\mathrm{c} \hat{c}^\dagger \hat{c}+ \omega_\mathrm{m} \hat{m}^\dagger \hat{m} + g(\hat{c}^\dagger \hat{m}+ \hat{c} \hat{m}^\dagger) +  \mathcal{K}\hat{m}^\dagger \hat{m} \hat{m}^\dagger \hat{m},
\end{equation}
where $\omega_\mathrm{c}$ is the angular resonance frequency of the cavity, $\hat{c}^\dagger (\hat{c})$ is the creation (annihilation) operator of the cavity mode, and $g$ is the magnon--photon coupling rate \cite{PhysRevLett.104.077202}. Regarding the magnon operators, $\langle \hat{m}^\dagger \hat{m} \rangle$ is the expectation value for the magnon harmonic oscillator population, and $\delta\hat{m}^\dagger \hat{m}$ is its fluctuation term. Using the mean field approximation, i.e., the condition that $\langle \hat{m}^\dagger \hat{m} \rangle \gg \delta\hat{m}^\dagger \hat{m}$ \cite{PhysRevB.94.224410}, the fluctuation term is negligible in the quadratic parts so that $\hat{m}^\dagger\hat{m} \hat{m}^\dagger \hat{m}=-\langle \hat{m}^\dagger \hat{m} \rangle^2+2\langle \hat{m}^\dagger \hat{m} \rangle \hat{m}^\dagger \hat{m}$, and we obtain
\begin{equation}\label{eq:Hamiltonian_full_MF}
    \hat{H}/\hbar = \omega_\mathrm{c} \hat{c}^\dagger \hat{c}+ (\omega_\mathrm{m} + 2\mathcal{K}\langle \hat{m}^\dagger \hat{m} \rangle) \hat{m}^\dagger \hat{m} + g(\hat{c}^\dagger \hat{m}+ \hat{c} \hat{m}^\dagger).
\end{equation}
The term $2\mathcal{K}\langle \hat{m}^\dagger \hat{m} \rangle$ is treated as a modulation of the magnon energy that depends on the magnon population. Reducing further the size $N$ of the system so that $\mathcal{K}\hat{m}^\dagger \hat{m} \hat{m}^\dagger \hat{m}$ becomes non-negligible even for small magnon numbers, thus not verifying $\langle \hat{m}^\dagger \hat{m} \rangle \gg \delta\hat{m}^\dagger \hat{m}$, would require a full quantum treatment of the Kerr nonlinearity to correctly describe the dynamics of the system, which is left for future work. From the Hamiltonian of \cref{eq:Hamiltonian_full_MF}, we derive the EOM following the Heisenberg-Langevin approach combined with input-output theory. We obtain
\begin{align}
    \begin{split}
        \Dot{c}(t) &= -\frac{i}{\hbar}\left [ c(t), \hat{H}\right] - \frac{\kappa_1^\mathrm{cpl}}{2}c(t) - \frac{\kappa_2^\mathrm{cpl}}{2}c(t) \\
        &+\sqrt{\kappa_1^\mathrm{cpl}}c_1^\mathrm{in}(t) + \sqrt{\kappa_2^\mathrm{cpl}} c_2^\mathrm{in}(t), \\
        \Dot{m}(t) & = - \frac{i}{\hbar} \left [m(t), \hat{H}\right]. 
    \end{split}
\end{align}
Here, $c_1^\mathrm{in}(t)$, $c_1^\mathrm{out}(t)$, $c_2^\mathrm{in}(t)$, $c_2^\mathrm{out}(t)$ are the input-output fields for port one and two, and $\kappa_1^\mathrm{cpl}$, $\kappa_2^\mathrm{cpl}$ are the associated coupling coefficients for the respective port, as represented schematically in \cref{fig:fig_setup}. Boundary conditions at the input-output ports give \cite{walls1994quantum}

\begin{align}
    \begin{split}\label{eq:BC_input_output}
        \sqrt{\kappa_1^\mathrm{cpl}}c(t) & = c_1^\mathrm{in}(t) + c_1^\mathrm{out}(t) , \\
        \sqrt{\kappa_2^\mathrm{cpl}}c(t) & = c_2^\mathrm{in}(t) + c_2^\mathrm{out}(t).
    \end{split}
\end{align}
Then we set
\begin{align}
    \begin{split}
        c(t) &= Ce^{-i\omega t},\\
        m(t) &= Me^{-i\omega t},\\
        c_1^\mathrm{in}(t) &= c_1^\mathrm{in}e^{-i\omega t},\\
        c_1^\mathrm{out}(t) &= c_1^\mathrm{out}e^{-i\omega t},\\
        c_2^\mathrm{in}(t) &= c_2^\mathrm{in}e^{-i\omega t},\\
        c_2^\mathrm{out}(t) &= c_2^\mathrm{out}e^{-i\omega t},
    \end{split}
\end{align}
where $C, M, c_1^\mathrm{in}, c_1^\mathrm{out}, c_2^\mathrm{in}, c_2^\mathrm{out}$ are time-independent amplitudes, and $\omega$ is the probing/driving angular frequency. Introducing the FWHM of internal cavity dissipation rate $\kappa_\mathrm{c}$, and both constant and magnon-number-dependent parts of the magnon dissipation rate, $\kappa_\mathrm{m}$ and $\kappa_\mathrm{m}'$, respectively, the EOM become
\begin{align}
    \begin{split}
         -i\omega C &= -i\omega_\mathrm{c} C - \frac{\kappa_\mathrm{tot}}{2}C - i g M \\
         & + \sqrt{\kappa_1^\mathrm{cpl}}c_1^\mathrm{in} +\sqrt{\kappa_2^\mathrm{cpl}}c_2^\mathrm{in},\\
        -i\omega M & = -i(\omega_\mathrm{m} + 2\mathcal{K}|M|^2)M \\
        & - \frac{\kappa_\mathrm{m}}{2}M - \frac{\kappa_\mathrm{m}'}{2}|M|^2M - igC.
    \end{split}
\end{align}
We assign $\kappa_\mathrm{tot}/(2\pi) = (\kappa_1^\mathrm{cpl} + \kappa_2^\mathrm{cpl} + \kappa_\mathrm{c})/(2\pi)$ to be the total cavity loss and set
\begin{align}
    \begin{split}
        A_1&=i\omega_\mathrm{c} -i\omega+\frac{\kappa_\mathrm{tot}}{2},\\
        C_\mathrm{in}&=\frac{1}{A_1}\left(\sqrt{\kappa_1^\mathrm{cpl}}c_1^\mathrm{in} +\sqrt{\kappa_2^\mathrm{cpl}} c_2^\mathrm{in}\right),
    \end{split}
\end{align}
to obtain
\begin{align}
    \begin{split}
        C &= -i\frac{g}{A_1}M+C_\mathrm{in},\\
        M &= \frac{igC_\mathrm{in}}{i(\omega-\omega_\mathrm{m}-2\mathcal{K}|M|^2)-\frac{\kappa_\mathrm{m}}{2}-\frac{\kappa_\mathrm{m}'}{2}|M|^2-\frac{g^2}{A_1}}.
        \label{eq:A_and_B}
    \end{split}
\end{align}
We first use the second line of this relation to obtain a cubic equation for the magnon population $|M|^2$
\begin{equation}\label{eq:cubic_M_sq}
    |M|^6A_5+|M|^4A_4+|M|^2A_3-A_2=0,
\end{equation}
with the real coefficients
\begin{align}\label{eq:cubic_M_sq_coeffs}
    \begin{split}
        A_5 &=\left(\delta_\mathrm{c}\frac{\kappa_\mathrm{m}'}{2}+\mathcal{K}\kappa_\mathrm{tot}\right)^2+\left(\frac{\kappa_\mathrm{tot}\kappa_\mathrm{m}'}{4}-2\mathcal{K}\delta_\mathrm{c}\right)^2,\\
        A_4 &= \mathcal{K}\delta_\mathrm{m}\kappa_\mathrm{tot}^2-4\mathcal{K}\delta_\mathrm{c}(g^2-\delta_\mathrm{c}\delta_\mathrm{m})+\delta_\mathrm{c}^2\kappa_\mathrm{m}\frac{\kappa_\mathrm{m}'}{2}\\
        &+\frac{\kappa_\mathrm{tot}^2\kappa_\mathrm{m}\kappa_\mathrm{m}'}{8}+\frac{\kappa_\mathrm{tot}\kappa_\mathrm{m}'g^2}{2},\\
        A_3 &= \left(\frac{\kappa_\mathrm{tot}} {2}\delta_\mathrm{m}+\delta_\mathrm{c}\frac{\kappa_\mathrm{m}}{2}\right)^2+\left(\delta_\mathrm{c}\delta_\mathrm{m}-\frac{\kappa_\mathrm{tot}\kappa_\mathrm{m}}{4}-g^2\right)^2,\\
        A_2 &= g^2|C_\mathrm{in}|^2\left(\frac{\kappa_\mathrm{tot}^2}{4}+\delta_\mathrm{c}^2\right).
    \end{split}
\end{align}
Here we have introduced the detuning terms for the cavity and magnon angular frequencies
\begin{align}
    \begin{split}
        \omega_\mathrm{c}-\omega &= \delta_\mathrm{c},\\
        \omega_\mathrm{m}-\omega &= \delta_\mathrm{m}.        
    \end{split}
\end{align}
After solving for the magnon population $|M|^2$ using the cubic relation, we can then obtain the photon population $|C|^2$ from \cref{eq:A_and_B}, which fully determines the dynamics of the hybrid system. 

The case of intermediate external field angles is more intricate, because the actual quantization angle $\theta_{\rm{M}}'$ that determines the parameters in \cref{eq:cubic_M_sq_coeffs} depends on the magnon population $|M|^2$ as well. Proceeding from \cref{eq:Hamiltonian_theta_Bog_third} within the mean field approximation leads to
\begin{equation}
    \begin{split}
        \op{H}_{\theta}^{(\rm{dir})}=&\hphantom{+}\hbar^{1/2}(u-v)(\op{m}+\hc{m})\\
        &\left[\gamma{}B_{\rm{ext}}\sin{(\theta_{\rm{M}}'-\theta)}\sqrt{\frac{S}{2}}+(2S)^{3/2}\mathcal{C}\cos{\theta_{\rm{M}}'}\sin{\theta_{\rm{M}}'}\right.\\
        &-(u^2-uv+v^2)\hbar|M|^2\\
        &\left.\left(\frac{\gamma{}B_{\rm{ext}}\sin{(\theta_{\rm{M}}'-\theta)}}{4\sqrt{2S}}+5\sqrt{\frac{S}{2}}\mathcal{C}\cos{\theta_{\rm{M}}'}\sin{\theta_{\rm{M}}'}\right)\right],
    \end{split}
\end{equation}
which is zero for
\begin{equation}\label{eq:Hamiltonian_equilibrium_dyn}
    \begin{split}
        &\gamma{}B_{\rm{ext}}\sin{(\theta_{\rm{M}}'-\theta)}\left[1-\frac{2\mathcal{R}-1}{2\sqrt{\mathcal{R}^2-1}}\frac{|M|^2}{2N}\right]\\
        &=-4\mathcal{C}S\cos{\theta_{\rm{M}}'}\sin{\theta_{\rm{M}}'}\left[1-\frac{2\mathcal{R}-1}{2\sqrt{\mathcal{R}^2-1}}\frac{5|M|^2}{2N}\right].
    \end{split}
\end{equation}
This new equilibrium condition on $\theta_{\rm{M}}'$ replaces \cref{eq:Hamiltonian_equilibrium} on $\theta_{\rm{M}}$.

\section{Simulation results}\label{sec:results}
Here we predict the IP and OOP Kerr coefficients expected for different materials, and show how they scale with magnon frequency and magnetic volume. We focus on YIG, Co$_{25}$Fe$_{75}$ (CoFe) and Permalloy (Ni$_{80}$Fe$_{20}$, NiFe), as they have been used in magnonics for a variety of purposes \cite{Serga_2010, Kruglyak_2010, Guo2023, Schoen2016, PhysRevLett.123.107701, PhysRevLett.123.107702, 10.1063/1.1656733, 10.1063/1.2197087, Zhao2016}. We then simulate the magnon and photon populations for a YIG thin film and relate them to the transmission of the cavity as relevant for experiments.

\subsection{Frequency and material dependence of Kerr coefficients}
\label{sec:freq_dep}

\begin{figure}[t]
\includegraphics[width=\linewidth, trim=0 0.25in 0 0]{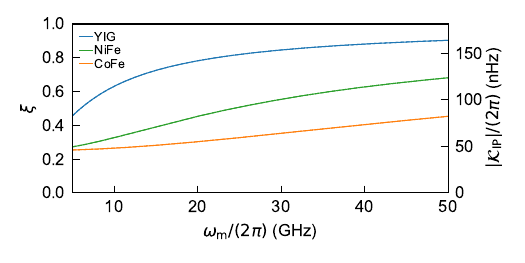}%
\caption{Ellipticity parameter $\xi$ (left axis) and magnitude of IP Kerr coefficient (right axis) calculated for YIG, CoFe, and NiFe from \cref{eq:xi_vs_omega_m,eq:Kerr_coeff_IP_simplified} as functions of $\omega_\mathrm{m}/(2\pi)$. The magnetic sample volume $V_\mathrm{m}$ is fixed to $\SI{3}{mm} \times \SI{3}{mm} \times \SI{100}{nm}$.}
\label{fig:fig1}
\end{figure}

For ferromagnetic materials, the analysis introduced in \cref{sec:theory} reveals that whereas $\mathcal{K}_\mathrm{OOP}$ solely depends on the magnetic volume $V_\mathrm{m}$ (or total number of spins $N$), dipolar effects related to the saturation magnetization $M_\mathrm{s}$ have an impact on $\mathcal{K}_\mathrm{IP}$ through the ellipticity parameter $\xi$ [see \cref{eq:Kerr_coeff_IP_simplified}]. \Cref{fig:fig1} shows $\xi$ and $|\mathcal{K}_\mathrm{IP}|$ as functions of $\omega_\mathrm{m}/(2\pi)$ calculated for YIG, CoFe, and NiFe, using \cref{eq:xi_vs_omega_m,eq:Kerr_coeff_IP_simplified}. We set the magnetic sample volume $V_\mathrm{m}=\SI{3}{mm} \times \SI{3}{mm} \times \SI{100}{nm}$, $\gamma/(2\pi)=\SI{28}{GHz/T}$, and the material parameters given by \Cref{tab:material_param}. The increase of $\xi$ versus frequency highlights that the magnetization precession becomes increasingly circular, as the Zeeman energy increases in comparison to the dipolar energy. Materials with larger $M_\mathrm{s}$ exhibit a reduced parameter $\xi$ at a given frequency, hence smaller $|\mathcal{K}_\mathrm{IP}|$ in the given frequency range. The parameter $\xi$ evolves between a lower limit of $1/4$ for lowest frequencies and largest $M_\mathrm{s}$, and an upper limit of unity for highest frequencies and smallest $M_\mathrm{s}$. Therefore, $\mathcal{K}_\mathrm{IP}$ saturates to the same value for the different materials when $\omega_\mathrm{m}$ increases indefinitely, reaching half the value of $\mathcal{K}_\mathrm{OOP}$ in
magnitude. For OOP field, $\mathcal{K}_\mathrm{OOP}/(2\pi)=\SI{363}{nHz}$ for the present volume, corresponding to an extended film, independent of $\omega_\mathrm{m}$ and $M_\mathrm{s}$.
\footnotesize
\begin{table}[t]
  \caption{Saturation magnetization, and ranges for Gilbert damping constant, IP and OOP inhomogeneous linewidth for YIG \cite{Hansen1974a}, CoFe \cite{Schoen2016} and NiFe \cite{10.1063/1.1656733, 10.1063/1.2197087, Zhao2016} for calculation of $\xi$ shown in \cref{fig:fig1} and of the figures of merit shown in \cref{fig:fig2}.}
  \hspace*{0pt}
  \begin{minipage}{\columnwidth}
  \begin{tabular*}{1\textwidth}{@{\extracolsep{\fill}}ccccc}
      \hline\hline
      Material &  $M_\mathrm{s}$ & $\alpha$  & $\Delta B_0^\mathrm{IP}$ & $\Delta B_0^\mathrm{OOP}$\\
      & $(\si{kA.m^{-1}})$  & $(10^{-3})$ & $(\SI{}{mT})$ & $(\SI{}{mT})$ \\
      \hline
      YIG & 196 & $0.05-0.1$ & $0.1-0.5$ & $0.5-1$ \\
      CoFe & 1990 &  $2-5$ & $0.5-5$ & $1-10$ \\
      NiFe & 796 &  $7-10$ & $0-5$ & $0-10$ \\
      \hline\hline
  \end{tabular*}
  \end{minipage}
  \label{tab:material_param}
\end{table}
\normalsize

Experimentally, the MKE has been demonstrated using bulk YIG spheres \cite{PhysRevB.94.224410, PhysRevLett.120.057202, PhysRevLett.129.123601, Shen2025}, extended thin films \cite{x4f8-pfj2} and patterned magnets \cite{zw18-26nw} as the magnonic system. The Kerr coefficient in spheres reaches about \SI{1}{nHz} at most, due to the lower bound for their volumes and small magnetocrystalline anisotropy. For extended thin films, the Kerr coefficient can be already three orders of magnitude higher due to strong shape anisotropy and smaller volumes, further reinforced in patterned magnets. Still, YIG spheres are widely used for inducing magnon Kerr nonlinearities due to their ultralow magnon dissipation rates, allowing for large magnon populations for a given input power. To evaluate the possibility of resolving the Kerr shift in experiments, the anharmonicity of magnonic systems should be compared taking into account both the magnitude of the Kerr coefficient and the magnon dissipation rate, i.e., $2\mathcal{K}/\kappa_\mathrm{m}$. Hence, we investigate how the normalized anharmonicity $2\mathcal{K}/\kappa_\mathrm{m}$ scales for different thin-film magnetic systems.

\begin{figure}[b]
\includegraphics[width=\linewidth, trim=0 0.25in 0 0]{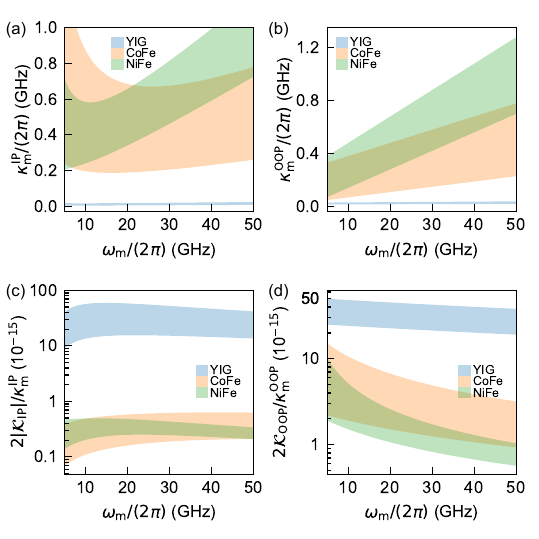}%
\caption{Calculated ranges of the magnon dissipation rate $\kappa_\mathrm{m}$ (FWHM) for YIG, CoFe, and NiFe in the (a) IP and (b) OOP cases, using the parameters of \cref{tab:material_param}. Corresponding normalized anharmonicities $2|\mathcal{K}|/\kappa_\mathrm{m}$ (c) for IP and (d) for OOP as functions of $\omega_\mathrm{m}/(2\pi)$. The magnetic sample volume $V_\mathrm{m}$ is fixed to $\SI{3}{mm} \times \SI{3}{mm} \times \SI{100}{nm}$.}
\label{fig:fig2}
\end{figure}

The magnon dissipation rate $\kappa_\mathrm{m}$ (FWHM) relates to the FWHM magnon resonance linewidth $\Delta B$ in field units according to
\begin{equation}
    \label{eq:freq_linewidth}
    \kappa_\mathrm{m} = \Delta B \frac{\partial \omega_\mathrm{m}}{\partial B_\mathrm{ext}},
\end{equation}
where $\partial \omega_\mathrm{m}/\partial B_\mathrm{ext}$ is the field-derivative of the Kittel relation. For IP field, it is
\begin{equation}
    \frac{\partial \omega_\mathrm{m}}{\partial B_\mathrm{ext}}=\gamma\frac{2B_\mathrm{ext}+B_\mathrm{eff}}{2\sqrt{B_\mathrm{ext}(B_\mathrm{ext}+B_\mathrm{eff})}},
\end{equation}
and for OOP field, $\partial \omega_\mathrm{m}/\partial B_\mathrm{ext} = \gamma $. The magnon resonance linewidth $\Delta B$ can be approximated as a linear function of frequency in a relevant window extending between $5$ and \SI{50}{GHz} \cite{PhysRevMaterials.4.024416} and is given by
\begin{equation}\label{eq:field_linewidth}
    \Delta B = \Delta B_0 + 2\alpha\omega_\mathrm{m}/\gamma,
\end{equation}
where $\Delta B_0$ is the linewidth at zero frequency due to inhomogeneous broadening and $\alpha$ is the Gilbert damping constant. \Cref{fig:fig2}(a) and (b) show the ranges of $\kappa_\mathrm{m}$ calculated for the three materials as functions of $\omega_\mathrm{m}/(2\pi)$ using the material parameters from \Cref{tab:material_param}, for IP and OOP fields, respectively. For IP field, there is a strong non-linear evolution of $\kappa_\mathrm{m}$ at lower frequencies, whereas for OOP field, $\kappa_\mathrm{m}$ is linear.

\Cref{fig:fig2}(c) and (d) show the normalized anharmonicity $2\mathcal{K}/\kappa_\mathrm{m}$ of the three systems as functions of $\omega_\mathrm{m}/(2\pi)$ for IP and OOP fields, respectively. For IP field, although the Kerr coefficient increases (\cref{fig:fig1}), the normalized anharmonicity decays at higher frequencies due to $\kappa_\mathrm{m}$. Moreover, $2|\mathcal{K}_\mathrm{IP}|/\kappa_\mathrm{m}^\mathrm{IP}$ peaks at a specific value determined by the minima in $\kappa_\mathrm{m}^\mathrm{IP}$ and $M_\mathrm{s}$ of the material. This defines a range in $\omega_\mathrm{m}/(2\pi)$ where the normalized anharmonicity is the strongest for a given material. With the present example values for YIG, it is near $\SI{15}{GHz}$, for CoFe close to $\SI{40}{GHz}$, and for NiFe near $\SI{10}{GHz}$. For OOP field, the normalized anharmonicity is the largest at the lowest resonance frequencies, where $\kappa_\mathrm{m}^\mathrm{OOP}$ is the smallest.

\subsection{Volume-dependence of anharmonicity}

\begin{figure}[t]
\includegraphics[width=\linewidth]{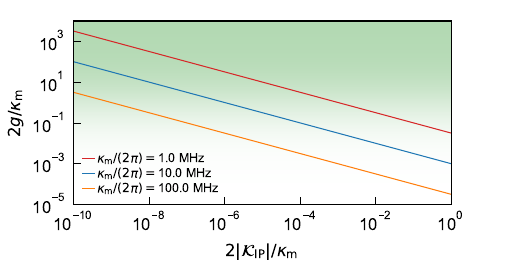}%
\caption{Ratio between coupling constant and magnon dissipation rate as a function of normalized anharmonicity for $g_0/(2\pi) = \SI{250}{Hz}$ and $\kappa_\mathrm{m}/(2\pi)=1.0,\ 10.0,\ \SI{100}{MHz}$. The gradient fill marks the transition from coherent ($2g\geq\kappa_\mathrm{m}$) to dissipative ($2g<\kappa_\mathrm{m}$) coupling.}
\label{fig:fig3}
\end{figure}

In this section, we investigate the volume dependence of the figures of anharmonicity for YIG in the IP configuration. Smaller magnetic volumes result in larger Kerr coefficients [\cref{eq:ip_Kerr}], and lithographed nanomagnets can have volumes easily as small as \SI{e-9}{mm^3} \cite{PhysRevLett.123.107701, PhysRevLett.123.107702, Guo2023}, significantly enhancing the anharmonicity of the magnonic system. However, decreasing the volume also reduces the magnon--photon coupling constant, which may result in poor sensitivity in experiments relying on the hybrid cavity--magnon modes. The magnon--photon coupling constant is $g=g_0\sqrt{N}$ where $g_0$ is the single-spin coupling strength of the microwave resonator mode and $N$ is the number of spins in the magnetic system. Using $n_\mathrm{s}=\SI{2e28}{m^{-3}}$ for the spin density in YIG and $g_0/(2\pi)=\SI{250}{Hz}$ as the single-spin coupling strength for a standard superconducting lumped-element resonator \cite{PhysRevLett.123.107702}, it is necessary to balance the increase in $2\mathcal{K}/\kappa_\mathrm{m}$ and the decrease in $2g/\kappa_\mathrm{m}$. \Cref{fig:fig3} shows $2g/\kappa_\mathrm{m}$ as a function of $2|\mathcal{K}_\mathrm{IP}|/\kappa_\mathrm{m}$
for different $\kappa_\mathrm{m}$ at $\omega_\mathrm{m}/(2\pi)=\SI{10}{GHz}$. Assuming a sufficiently low cavity loss ($g\gg \kappa_\mathrm{c}$), the gradient shading marks the transition from $2g\geq\kappa_\mathrm{m}$ to $2g<\kappa_\mathrm{m}$, i.e., highlighting where the system crosses from the coherent-interaction regime into the dissipation-dominated regime, as the normalized anharmonicity is increased. Decreasing the magnon dissipation rate increases both the coherence of the cavity--magnon interaction and the normalized anharmonicity, making it an important material parameter for inducing strong magnon nonlinearities. To observe magnon antibunching \cite{PhysRevB.102.100402}, which requires $2|\mathcal{K}| \sim \kappa_\mathrm{m}$, one has to reduce the number of spins to $N\approx 4000$, e.g., a YIG platelet with $V_{\rm{m}}=$~\qtyproduct[product-units=power]{10x10x2}{\nano\meter}, for $\kappa_\mathrm{m}/(2\pi)=\SI{1}{MHz}$. At this $N$, a single-spin coupling strength $g_0/(2\pi)\approx\SI{8}{kHz}$ is required to remain within the strong coupling regime ($2g \geq \kappa_\mathrm{m}$) \cite{pscherer2026superconductingparallelplateresonatorsdetection}. 

\subsection{Angle-dependence of magnon Kerr coefficient}\label{sec:kerr_angle_simulation}
\begin{figure}[ht]
\includegraphics[width=\linewidth]{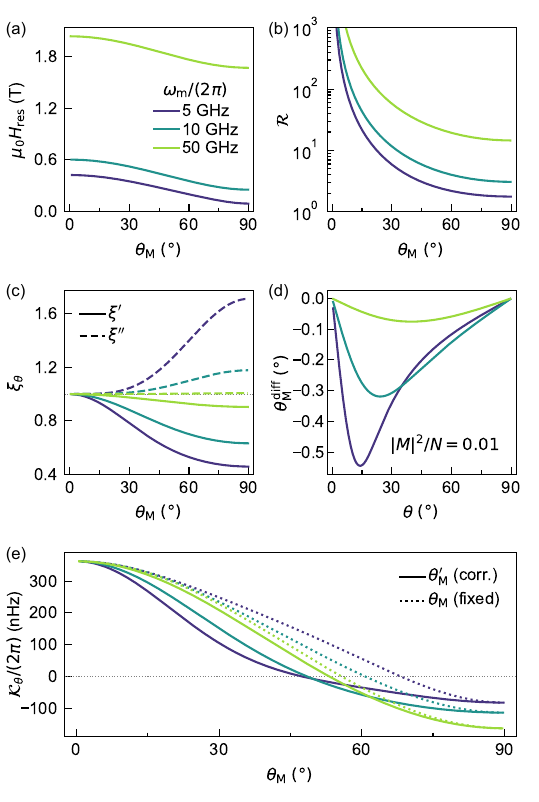}%
\caption{(a) Ferromagnetic resonance field $\mu_0 H_\mathrm{res}$, (b) $\mathcal{R}$ based on \cref{eq:R_theta}, (c) $\xi'$ and $\xi''$ based on \cref{eq:xip_xipp}, (d) $\theta_\mathrm{M}^\mathrm{diff}=\theta_\mathrm{M}'-\theta_\mathrm{M}$, and (e) angle-dependent magnon Kerr coefficient $\mathcal{K}_\theta/(2\pi)$, all computed for YIG at \SI{0}{K} and at $\omega_\mathrm{m}/(2\pi)=5$, 10, \SI{50}{GHz} versus $\theta_\mathrm{M}$ (a--c, e) and $\theta$ (d), using $M_\mathrm{s}=\SI{196}{kA/m}$, $V_\mathrm{m}=\SI{3}{mm} \times \SI{3}{mm} \times \SI{100}{nm}$, and $\gamma/(2\pi)=\SI{28}{GHz/T}$. The reoriented angle $\theta_\mathrm{M}'$ in (d) is obtained from \cref{eq:Hamiltonian_equilibrium_dyn} for $|M|^2/N=0.01$ at the resonance field of (a). In (e), dotted lines show $\mathcal{K}_\theta/(2\pi)$ from \cref{eq:theta_Kerr} at a fixed magnetization angle, and solid lines show $\mathcal{K}_{\rm{tot}}/(2\pi)$ that includes the additional effect of population-dependent reorientation of the magnetization given by \cref{eq:Hamiltonian_equilibrium_dyn}. The parameter $\mathcal{K}_{\rm{tot}}/(2\pi)$ corresponds to half the slope of the total frequency shift versus $|M|^2$, evaluated in the limit $|M|^2\rightarrow0$, where the shift is proportional to the magnon population. In all panels, colors denote $\omega_\mathrm{m}/(2\pi)$ as indicated in (a); the legends of (c) and (e) only identify the line styles, i.e., solid for $\xi'$ and dashed for $\xi''$ in (c), and solid for the reoriented and dotted for the fixed magnetization angle in (e).}
\label{fig:fig_Kerr_angle}
\end{figure}

In this section, we study the magnetization angle-dependence of the magnon Kerr coefficient $\mathcal{K}_\theta$ for a YIG thin film in the \SI{0}{K} limit. The volume $V_\mathrm{m}$ is fixed to $\SI{3}{mm} \times \SI{3}{mm} \times \SI{100}{nm}$ and we set the resonance frequency $\omega_\mathrm{m}/(2\pi)=5, 10, \SI{50}{GHz}$. We start by calculating the field-angle dependence of the resonance field $\mu_0 H_\mathrm{res}$ at this fixed frequency using \cref{eq:Hamiltonian_theta_Bog_harm}, where the equilibrium magnetization angle $\theta_\mathrm{M}$ is obtained from \cref{eq:Hamiltonian_equilibrium}. The resulting magnetization angle-dependence of $\mu_0 H_\mathrm{res}$ is shown in \cref{fig:fig_Kerr_angle}(a), evolving continuously between the OOP ($\theta_\mathrm{M}=\SI{0}{\degree}$) and IP ($\theta_\mathrm{M}=\SI{90}{\degree}$) magnetic configurations.

\Cref{fig:fig_Kerr_angle}(b) shows $\mathcal{R}$ versus $\theta_\mathrm{M}$ based on \cref{eq:R_theta}, using the values of $\mu_0 H_\mathrm{res}$ from \cref{fig:fig_Kerr_angle}(a). As expected, $\mathcal{R}>1$ for all angles and diverges as $\theta_\mathrm{M}$ approaches $\SI{0}{\degree}$. Using this, we obtain $\xi'$ and $\xi''$ versus $\theta_\mathrm{M}$ from \cref{eq:xip_xipp}, shown in \cref{fig:fig_Kerr_angle}(c). Here, $\xi''>1$ and $\xi'<1$, and $\xi'\rightarrow \xi$ given by \cref{eq:xi} when $\theta_\mathrm{M}\rightarrow\SI{90}{\degree}$. Moreover, as $\omega_\mathrm{m}$ increases, both $\xi'$ and $\xi''$ approach unity because the magnetic precession becomes increasingly circular, as the external field grows much larger compared to the anisotropy field.

To see the effects of large magnon populations on the equilibrium magnetization angle at intermediate field angles $\theta$, we compare in \cref{fig:fig_Kerr_angle}(d) the difference between $\theta_\mathrm{M}$ obtained from \cref{eq:Hamiltonian_equilibrium} and $\theta_\mathrm{M}'$ obtained from \cref{eq:Hamiltonian_equilibrium_dyn} with $|M|^2/N=0.01$. The difference $\theta_\mathrm{M}^\mathrm{diff}=\theta_\mathrm{M}'-\theta_\mathrm{M}$ is larger in magnitude at smaller frequencies, and also displays an extremum at $\theta\leq\SI{45}{\degree}$ that depends on $\omega_\mathrm{m}$. The largest $|\theta_\mathrm{M}^\mathrm{diff}|$ is in the range of $\SI{0.5}{\degree}$, and despite this angle shift being comparatively small, it leads to a noticeable correction of the resonance frequency and hence of the total nonlinearity $\mathcal{K}_{\rm{tot}}/(2\pi)$. \Cref{fig:fig_Kerr_angle}(e) presents $\mathcal{K}_\theta/(2\pi)$ versus $\theta_\mathrm{M}$ obtained from \cref{eq:theta_Kerr} at a fixed magnetization angle (dotted lines), and compares it with the coefficient $\mathcal{K}_{\rm{tot}}/(2\pi)$ that accounts for the population-dependent reorientation of the magnetization, $\theta_\mathrm{M} \rightarrow \theta_\mathrm{M}'$, given by \cref{eq:Hamiltonian_equilibrium_dyn} (solid lines), calculated for this YIG thin film. This highlights the continuous change of $\mathcal{K}_\theta$ between $\mathcal{K}_\mathrm{OOP}$ and $\mathcal{K}_\mathrm{IP}$, and the cancellation or sign change of the MKE near the magic angle, which has been observed experimentally in our related work \cite{x4f8-pfj2}. Moreover, the dynamical reorientation of the magnetization angle shifts the $\mathcal{K}_{\rm{tot}}=0$ points to lower $\theta_\mathrm{M}$, and this change is larger for lower $\omega_\mathrm{m}$, as can also be guessed from \cref{fig:fig_Kerr_angle}(d). At higher frequencies, $\mathcal{K}_{\rm{tot}}$ closely follows $\mathcal{K}_\theta$, with a minimized impact of the dynamical reorientation.

\subsection{Power-dependence of magnon-polariton frequency shifts}
In this section, we study the evolution of the photon and magnon populations as functions of input power in a hybrid cavity magnonics system with YIG, at $\omega_\mathrm{c}/(2\pi)=\omega_\mathrm{m}/(2\pi)=\SI{10}{GHz}$. We use \cref{eq:A_and_B} to obtain $|C|^2$ and $|M|^2$, which correspond to photon and magnon populations, respectively. We set $g/(2\pi) = \SI{30}{MHz}$, $\kappa_\mathrm{tot}/(2\pi)=\SI{20}{MHz}$ \cite{1mc9-k683} and $\kappa_\mathrm{m}/(2\pi)=\SI{5}{MHz}$. Since $g>\kappa_\mathrm{tot}/2,\kappa_\mathrm{m}/2$, the system is in the strongly coupled regime with cooperativity $4g^2/(\kappa_\mathrm{tot}\kappa_\mathrm{m})=36$. We also set $\kappa_1^\mathrm{cpl}=\kappa_2^\mathrm{cpl}=\kappa_\mathrm{tot}/4$ such that the cavity resonator achieves critical coupling with the input and output ports. The volume of the magnetic sample is the same as in \cref{sec:freq_dep}.

\begin{figure}[t]
\includegraphics[width=\linewidth,trim=0 .25in 0 0]{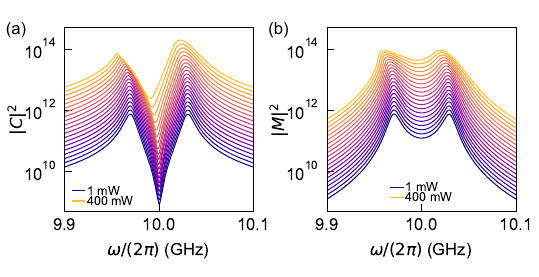}%
\caption{Population of (a) photons and (b) magnons as functions of drive frequency $\omega/(2\pi)$ for drive powers from \SI{1}{} to \SI{400}{mW}, logarithmically spaced, at $\omega_\mathrm{c}/(2\pi)=\omega_\mathrm{m}/(2\pi)=\SI{10}{GHz}$, field applied in the film plane (IP).}
\label{fig:fig4}
\end{figure}

Considering IP magnetization first, $\mathcal{K}_\mathrm{IP}/(2\pi) = \SI{-115}{nHz}$ at $\omega_\mathrm{m}/(2\pi)=\SI{10}{GHz}$, and we set the nonlinear magnon damping coefficient $\kappa_\mathrm{m}'/(2\pi)=\SI{200}{nHz}$ \cite{x4f8-pfj2}. \Cref{fig:fig4}(a) and (b) show the drive angular frequency dependence of $|C|^2$ and $|M|^2$ at different input powers. The two peaks in the spectra correspond to the lower and upper polariton (LP and UP) branches of the magnon--photon coupled system. Both $|C|^2$ and $|M|^2$ show the increase in their populations for higher powers. Moreover, the spectra also show the shifts of the polaritons as the population increases. The sign of $\mathcal{K}$ determines the shift direction for the polaritonic branches, where in the case of $\mathcal{K}<0$, such as here, the shift occurs mainly for the LP towards lower frequencies, whereas at $\mathcal{K}>0$, it would occur mainly for the UP towards higher frequencies. 

\begin{figure}[t]
\includegraphics[width=\linewidth,trim=0 .25in 0 0]{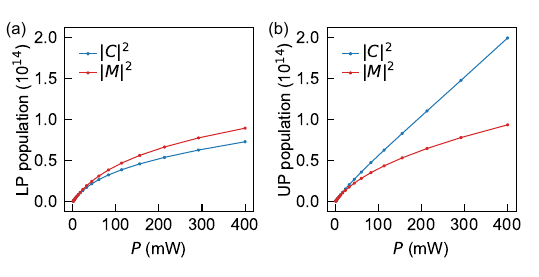}%
\caption{Photon and magnon populations for the (a) lower and (b) upper polaritons as functions of the input power, field applied in the film plane (IP), at $\omega_\mathrm{c}/(2\pi)=\omega_\mathrm{m}/(2\pi)=\SI{10}{GHz}$. For each power, $|C|^2$ and $|M|^2$ are taken at their respective maxima over the
drive frequency $\omega$ within each polariton branch.}
\label{fig:fig5}
\end{figure}

The input power dependence of the magnon and photon populations for the lower and upper polaritons are shown in \cref{fig:fig5}, for the IP case again, with $\mathcal{K}_\mathrm{IP}/(2\pi) = \SI{-115}{nHz}$. We see that as the power increases, the maximum magnon population in the LP [\cref{fig:fig5}(a)] becomes larger than its maximum photon population, which also results in a more pronounced frequency shift of the LP mode as the mode becomes more magnon-like. In contrast, the photon population in the UP [\cref{fig:fig5}(b)] becomes larger than its magnon population, as the UP becomes more photon-like with the Kerr shift. Hence, the photon population in the UP increases linearly with input power, as the magnon character becomes smaller. In the case of a positive $\mathcal{K}$, the behaviors for LP and UP would be opposite.

Considering now the same YIG film parameters, but in the OOP magnetized configuration for simplicity, the magnon Kerr coefficient $\mathcal{K}_\mathrm{OOP}/(2\pi)=\SI{363}{nHz}$ is larger and the power-dependent dissipation is suppressed, $\kappa_\mathrm{m}'=0$. We can derive an approximate formula for the microwave power $P$ required to obtain a magnon Kerr shift in the order of the magnon resonance linewidth. This regime of resolution of the Kerr shift requires a magnon population
\begin{equation}\label{eq:magnon_population_linewidth}
    |M|^2>\frac{\kappa_\mathrm{m}}{2\mathcal{K}}.
\end{equation}
To relate the magnon population to the input microwave power, we start from Eq.~\eqref{eq:A_and_B} with $\kappa_\mathrm{m}'=0$ and $\omega_\mathrm{c}=\omega_\mathrm{m}$, and tune $\omega$ to the upper magnon polariton, such that $\omega-\omega_\mathrm{m}-2\mathcal{K}|M|^2\approx g$, which also means that $\omega-\omega_\mathrm{c}\approx g+2\mathcal{K}|M|^2$. Then, we obtain
\begin{equation}
\begin{split}
    \hbar\omega|M|^2\Bigg[&\left(2\mathcal{K}|M|^2g-\frac{\kappa_\mathrm{tot}\kappa_\mathrm{m}}{4}\right)^2\\
    &+\left(g\frac{\kappa_\mathrm{tot}+\kappa_\mathrm{m}}{2}+\mathcal{K}\kappa_\mathrm{m}|M|^2\right)^2\Bigg]
    =g^2\kappa_1^\mathrm{cpl}P,
\end{split}
\label{eq:M2_vs_P}
\end{equation}
which, for the magnon population of Eq.~\eqref{eq:magnon_population_linewidth}, results in the power
\begin{equation}
    P>\frac{\kappa_\mathrm{m}}{2\mathcal{K}}\hbar \omega \frac{\left(g\kappa_\mathrm{m}-\frac{\kappa_\mathrm{tot}\kappa_\mathrm{m}}{4}\right)^2+\left(g\frac{\kappa_\mathrm{tot}+\kappa_\mathrm{m}}{2}+\frac{\kappa_\mathrm{m}^2}{2}\right)^2}{g^2 \kappa_1^\mathrm{cpl}}.
    \label{eq:P_linewidth}
\end{equation}
This relation identifies powers necessary for inducing a magnon Kerr shift in the order of the magnon resonance linewidth. For the YIG film in the OOP configuration considered here, it yields $P\gtrapprox \SI{11}{mW}$.

If instead of a YIG thin film, a YIG sphere of the same volume is considered, the smaller magnetocrystalline anisotropy would result in a correspondingly smaller $|\mathcal{K}|$, by one to two orders of magnitude (see \cref{fig:fig_spheres}). Since the magnon Kerr shift is proportional to $2\mathcal{K}|M|^2$ and $|M|^2$ is overall proportional to the input microwave power, the Kerr shift scales as $\mathcal{K}P$. To obtain the same Kerr shift, the required input power scales as $\sim1/|\mathcal{K}|$. Therefore, the enhancement of the Kerr coefficient by one to two orders of magnitude in thin films compared to spheres implies that the same nonlinear frequency shift can be obtained at input powers reduced by one to two orders of magnitude, substantially reducing the requirements to enter the nonlinear regime and observe foldover.

\subsection{Transmission via scattering matrix terms}
\begin{figure}[t]
\includegraphics[width=\linewidth,trim=0 .25in 0 0]{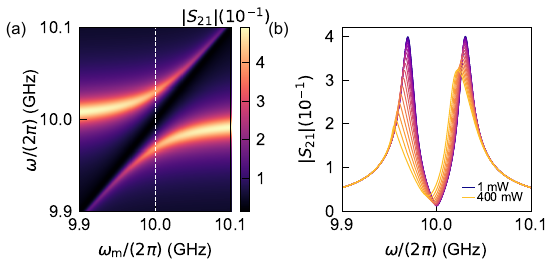}%
\caption{Amplitude of $S_{21}$ versus (a) drive and magnon frequencies at $P=\SI{1}{mW}$, and versus (b) drive frequency at input powers from \SI{1}{} to \SI{400}{mW}, logarithmically spaced, at $\omega_\mathrm{m}/(2\pi)=\SI{10}{GHz}$, i.e., along the dashed vertical line in (a). The field is applied in the film plane (IP).}
\label{fig:fig6}
\end{figure}

A two-port cavity resonator loaded with a ferromagnetic sample is commonly employed for studying magnon dynamics in the ferromagnet. The photon population in the cavity is detected by transmission between the two ports, and the MKE can be probed as frequency shifts of the magnon--polaritons in the output spectra \cite{PhysRevB.94.224410, PhysRevLett.120.057202, x4f8-pfj2}. The microwave transmission is often characterized by the $S_\mathrm{21}$ scattering matrix term, which is given by
\begin{equation}\label{eq:S21}
    S_\mathrm{21} = \frac{c_2^\mathrm{out}}{c_1^\mathrm{in}} = \frac{\sqrt{\kappa_2^\mathrm{cpl}}}{\sqrt{P/(\hbar\omega)}}C,
\end{equation}
assuming that $c_2^\mathrm{in}=0$. \Cref{fig:fig6}(a) shows a map of $|S_{21}|$ as a function of $\omega/(2\pi)$ and $\omega_\mathrm{m}/(2\pi)$ at $P=\SI{1}{mW}$, still in the linear regime,  using the simulation parameters from above for the YIG film magnetized in-the-plane. The map features the avoided crossing between the cavity and Kittel modes, associated with the strong coupling in the system. 

\Cref{fig:fig6}(b) shows a cross cut of the $|S_{21}|$ spectra at $\omega_\mathrm{m}/(2\pi)=\SI{10}{GHz}$, i.e., when the magnon mode overlaps with the cavity mode [dashed vertical line in \cref{fig:fig6}(a)]. The increase in power shifts the LP transmission branch similar to that in \cref{fig:fig4}, with the shift direction determined by the sign of $\mathcal{K}$, and consistent with recent experimental observations \cite{x4f8-pfj2}. Hence, the present modeling of the MKE and derivation of the EOM gives access to all the parameters that are relevant to nonlinear magnetization dynamics in experiments.

\subsection{Bistable solutions to the equations of motion}
The results presented so far were obtained in the regime where the cubic equation in $|M|^2$ [\cref{eq:cubic_M_sq}] only had a single root. In this section, we investigate the bistability regime in the EOM where \cref{eq:cubic_M_sq} can have multiple (up to three) roots. We investigate the same YIG film parameters but consider the OOP magnetized configuration, with larger magnon Kerr coefficient [$\mathcal{K}_\mathrm{OOP}/(2\pi)=\SI{363}{nHz}$] and suppressed power-dependent dissipation rate $\kappa_\mathrm{m}'$. This increases the normalized anharmonicity of the YIG system such that bistable regimes can be realized at moderate excitation powers.

\begin{figure}[t]
\includegraphics[width=\linewidth]{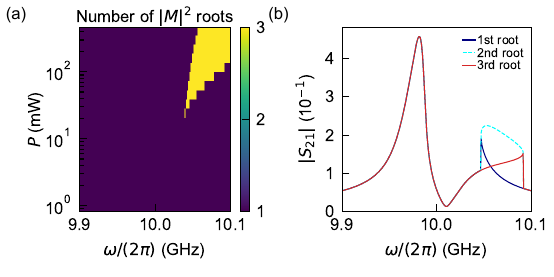}%
\caption{(a) Number of roots for $|M|^2$ in \cref{eq:cubic_M_sq} versus microwave power and microwave frequency, for our YIG film magnetized in the OOP configuration. (b) Transmission parameter $|S_{21}|$ versus microwave frequency at $\omega_\mathrm{c}/(2\pi)=\omega_\mathrm{m}/(2\pi)=\SI{10}{GHz}$ and $P=\SI{113}{mW}$ according to \cref{eq:S21} in the bistability regime using the three roots from \cref{eq:cubic_M_sq}.}
\label{fig:fig10}
\end{figure}

\Cref{fig:fig10}(a) shows the number of solutions to \cref{eq:cubic_M_sq} versus microwave frequency and power at $\omega_\mathrm{c}/(2\pi)=\omega_\mathrm{m}/(2\pi)=\SI{10}{GHz}$. The bistability regime with three roots in $|M|^2$ occurs in the upper right part of the map, for the UP, because of the magnon-like character of the UP at higher powers. Conversely, in the IP configuration with $\mathcal{K}_\mathrm{IP}<0$, the bistable regime would be the upper left quadrant, for the LP. \Cref{fig:fig10}(b) shows microwave transmission $|S_{21}|$ calculated using the three roots in $|M|^2$ versus microwave frequency, at $P=\SI{113}{mW}$. The solutions of $S_{21}$ using the first (smallest) and third (largest) root in $|M|^2$ are stable and have been observed experimentally \cite{PhysRevB.98.174423}, whereas the solution using the second root is unstable as the result of perturbations. Hence, EOM can also render the magnetization dynamics in the bistability regime.

\section{Conclusions}\label{sec:conclusions}

In conclusion, we have modeled the MKE for a thin film magnetized OOP, IP, and at an intermediate angle. For IP and intermediate-angle magnetizations, the non-collinearity between the external field and the anisotropy field directions gives rise to a factor related to precession ellipticity in the expression of the Kerr coefficient. We also introduce the correction to the Kerr coefficient that originates from the dynamical reorientation of the quantization axis of the magnetization at large excitation powers. We have shown how the Kerr coefficient and the anharmonicity scale with magnon frequency in extended thin films of different magnetic materials and in nanomagnets. The predicted figures of anharmonicity quantify the necessary trade-off between Kerr coefficient and magnon--photon coupling strength. Reducing the resonance linewidth of the magnon part and increasing the single-spin coupling constant of the photon part will lower the magnetic volume required for strong coupling, a necessary step for inducing stronger magnon Kerr nonlinearities in experiments. Analyzing the dependence on input power of both magnon and photon populations reveals the nonlinear magnon- and photon-like behaviors of the magnon--polariton branches, and their character varying with the applied power due to the detuning induced by the nonlinear Kerr shift, as well as the bistability regime. Finally, we derived the scattering matrix term $S_{21}$ to relate these populations and all system parameters to experimentally accessible quantities. These properties can be later used for tripartite entangled systems \cite{PhysRevLett.121.203601, PhysRevResearch.1.023021} or for parametric pumping of magnons \cite{Mukhopadhyay2022}.

\section*{Acknowledgments}

This work was supported by the Swiss National Science Foundation (Grant No.~10004526). We acknowledge CNRS Physics for supporting our work through an international research project (IRP 2025 CORMORANT). H.M.~acknowledges support from JSPS Postdoctoral Fellowship (Grant No.~23KJ1159). H.W.~acknowledges the support of the China Scholarship Council (CSC, Grant No.~202206020091). R.S.~acknowledges funding by the Deutsche Forschungsgemeinschaft via the SFB 1432, Project No. 425217212. W.L.~acknowledges the support of the ETH Zurich Postdoctoral Fellowship Program (21-1 FEL-48) and from a government grant managed by the Agence Nationale de la Recherche as part of the France 2030 program, with reference ANR-24-EXSP-0005 (“MAGNON-BRAQET”).

\section*{Data Availability}
The data that support the findings of this article are openly available \cite{Petrosyan2026data_MKE_theory}.

\clearpage
%

\appendix
\section{Operator formulas for magnons in the Holstein-Primakoff approximation}
\label{appendix:formulas}
In this Appendix we gather some useful formulas involved in solving and simplifying the Hamiltonians provided in \cref{sec:hamiltonian} and \cref{sec:eom}.

\subsection{Squares and products of spin operator}
Using the $[\op{b},\hc{b}]$ commutator, squaring the operators yields
\begin{align}
\label{eq:HP_Sx2}
    \begin{split}
        \op{S}_{x}^2=\frac{S}{2}
        &\left[\op{b}\op{b}+\hc{b}\hc{b}+\op{b}\hc{b}+\hc{b}\op{b}-\frac{\hc{b}\op{b}\hc{b}\op{b}}{S}\right.\\
        &\hphantom{[}\left.-\frac{\hc{b}\op{b}\op{b}\op{b}+\op{b}\hc{b}\op{b}\op{b}+\hc{b}\hc{b}\op{b}\hc{b}+\hc{b}\hc{b}\hc{b}\op{b}}{4S}\right],
    \end{split}
\end{align}
\begin{align}
\label{eq:HP_Sy2}
    \begin{split}
         \op{S}_{y}^2=\frac{S}{2}
        &\left[-\op{b}\op{b}-\hc{b}\hc{b}+\op{b}\hc{b}+\hc{b}\op{b}-\frac{\hc{b}\op{b}\hc{b}\op{b}}{S}\right.\\
        &\hphantom{[}\left.+\frac{\hc{b}\op{b}\op{b}\op{b}+\op{b}\hc{b}\op{b}\op{b}+\hc{b}\hc{b}\op{b}\hc{b}+\hc{b}\hc{b}\hc{b}\op{b}}{4S}\right],
    \end{split}
\end{align}
\begin{align}
\label{eq:HP_Sz2}
    \begin{split}
        \op{S}_{z}^2=S^2&-2S\hc{b}\op{b}+\hc{b}\op{b}\hc{b}\op{b}.
    \end{split}
\end{align}

Combining \cref{eq:HP_Sx2,eq:HP_Sy2} provides
\begin{align}
    \begin{split}
        \label{eq:HP_Sx2Sy2}
        &\op{S}_{x}^2-\op{S}_{y}^2\\&=S\left[
        \op{b}\op{b}
        +\hc{b}\hc{b}
        -\frac{\hc{b}\op{b}\op{b}\op{b}+\op{b}\hc{b}\op{b}\op{b}+\hc{b}\hc{b}\op{b}\hc{b}+\hc{b}\hc{b}\hc{b}\op{b}}{4S}
        \right].
    \end{split}
\end{align}

\subsection{Expansion of transformed magnon operators}

The Bogoliubov transformation of the creation and annihilation operators $\hc{b},\op{b}$ into $\hc{m},\op{m}$ is defined as 
\begin{equation}
    \hc{b}/\sqrt{\hbar}=u\hc{m}-v\op{m},\ \op{b}/\sqrt{\hbar}=u\op{m}-v\hc{m}.
\end{equation}
Preserving the canonical $[\op{b},\hc{b}]=\hbar$ into $[\op{m},\hc{m}]=1$ requires $u^2-v^2=1$. We note $\op{S}_1=\op{m}\hc{m}+\hc{m}\op{m}$ and $\op{S}_2=\hc{m}\hc{m}+\op{m}\op{m}$. Then, terms of third order are
\begin{equation}
    \label{eq:anharm_third_1}
    \begin{split}
    (\hc{b}\op{b}\op{b}+\hc{b}\hc{b}\op{b})/\hbar^{3/2}=&
    \hphantom{-}u^3(\hc{m}\op{m}\op{m}+\hc{m}\hc{m}\op{m})\\
    &-v^3(\op{m}\hc{m}\hc{m}+\op{m}\op{m}\hc{m})\\
    &+uv(v-u)(\hc{m}\hc{m}\hc{m}+\op{m}\op{m}\op{m})\\
    &-u^2v(\hc{m}\op{S}_1+\op{S}_1\op{m})\\&+uv^2(\op{m}\op{S}_1+\op{S}_1\hc{m})
    \end{split}
\end{equation}
and
\begin{equation}
    \label{eq:anharm_third_2}
    \begin{split}
    (\hc{b}\op{b}\hc{b}+\op{b}\hc{b}\op{b})/\hbar^{3/2}=\hphantom{+}
    (u^3-v^3)(\hc{m}\op{m}\hc{m}+\op{m}\hc{m}\op{m})\\
    +uv(v-u)(\hc{m}\hc{m}\hc{m}+\op{m}\op{m}\op{m}\\+\hc{m}\op{m}\op{m}+\op{m}\op{m}\hc{m}+\op{m}\hc{m}\hc{m}+\hc{m}\hc{m}\op{m})
    \end{split}
\end{equation}
while terms of fourth order are
\begin{equation}
    \label{eq:anharm_1}
    \begin{split}
        \hc{b}\op{b}\hc{b}\op{b}/\hbar^2&=\left(u^2\hc{m}\op{m}+v^2\op{m}\hc{m}-uv\op{S}_2\right)^2\\
        &=u^4\hc{m}\op{m}\hc{m}\op{m}+v^4\op{m}\hc{m}\op{m}\hc{m}\\&+u^2v^2(\hc{m}\op{m}\op{m}\hc{m}+\op{m}\hc{m}\hc{m}\op{m})\\
        &\hphantom{=\ }-u^3v(\hc{m}\op{m}\op{S}_2+\op{S}_2\hc{m}\op{m})\\
        &\hphantom{=\ }-uv^3(\op{m}\hc{m}\op{S}_2+\op{S}_2\op{m}\hc{m})
       +u^2v^2\op{S}_2^2
    \end{split}
\end{equation}
and
\begin{align}
    \label{eq:anharm_2}
    \begin{split}
        (\hc{b}\op{b}+\op{b}\hc{b})\op{b}\op{b}/\hbar^2&=\left[(u^2+v^2)\op{S}_1-2uv\op{S}_2\right]\\
        &\hphantom{=\ }\left[u^2\op{m}\op{m}+v^2\hc{m}\hc{m}-uv\op{S}_1\right]\\
    \end{split}\\
    \label{eq:anharm_3}
    \begin{split}
        \hc{b}\hc{b}(\hc{b}\op{b}+\op{b}\hc{b})/\hbar^2&=\left[u^2\hc{m}\hc{m}+v^2\op{m}\op{m}-uv\op{S}_1\right]\\
        &\hphantom{=\ }\left[(u^2+v^2)\op{S}_1-2uv\op{S}_2\right]\\
    \end{split}
\end{align}
so that
\begin{equation}
    \label{eq:anharm_4}
    \begin{split}           
        &(\hc{b}\op{b}+\op{b}\hc{b})\op{b}\op{b}/\hbar^2+\hc{b}\hc{b}(\hc{b}\op{b}+\op{b}\hc{b})/\hbar^2=\\
        &\hphantom{+\ }u^4(\op{S}_1\op{m}\op{m}+\hc{m}\hc{m}\op{S}_1)+v^4(\op{S}_1\hc{m}\hc{m}+\op{m}\op{m}\op{S}_1)\\
        &+3u^2v^2(\op{S}_1\op{S}_2+\op{S}_2\op{S}_1)-2uv(u^2+v^2)(\op{S}_1)^2\\
        &-2u^3v(\op{S}_2\op{m}\op{m}+\hc{m}\hc{m}\op{S}_2)-2uv^3(\op{S}_2\hc{m}\hc{m}+\op{m}\op{m}\op{S}_2).
    \end{split}
\end{equation}

Beyond $u^2-v^2=1$, some terms appearing are 
\begin{align}
    u^2+v^2=2uv\mathcal{R}&=\frac{\mathcal{R}}{\sqrt{\mathcal{R}^2-1}};\\
    u^2-uv+v^2&=\frac{2\mathcal{R}-1}{2\sqrt{\mathcal{R}^2-1}};\\
    (u-v)^2(u^2-uv+v^2)&=1-\frac{3}{2(\mathcal{R}+1)};\\
    u^2v^2&=\frac{1}{4(\mathcal{R}^2-1)}
\end{align}

\subsection{Operators in the rotated frame of reference}
\label{sec:rotated_HP}

After redefinition of the quantization axis along $\theta_{\rm{M}}$ with new operators
\begin{align}
    \op{S}_m&=\sin{\theta_{\rm{M}}}\op{S}_x+\cos{\theta_{\rm{M}}}\op{S}_z,\\
    \op{S}_n&=\cos{\theta_{\rm{M}}}\op{S}_x-\sin{\theta_{\rm{M}}}\op{S}_z,
\end{align}
the magnon operators are
\begin{align}
    \label{eq:HP_Sn}
    \op{S}_{n}&=\sqrt{\frac{S}{2}}\left[\op{b}-\frac{\hc{b}\op{b}\op{b}}{4S}+\hc{b}-\frac{\hc{b}\hc{b}\op{b}}{4S}\right],\\
    \label{eq:HP_Syp}
    \op{S}_{y}&=-i\sqrt{\frac{S}{2}}\left[\op{b}-\frac{\hc{b}\op{b}\op{b}}{4S}-\hc{b}+\frac{\hc{b}\hc{b}\op{b}}{4S}\right],\\
    \label{eq:HP_Sm}
    \op{S}_{m}&=S-\hc{b}\op{b},
\end{align}
so that, in particular, 
\begin{equation}
\label{eq:HP_SmSn_SnSm}
\begin{split}
\op{S}_m\op{S}_n+\op{S}_n\op{S}_m=&\sqrt{\frac{S}{2}}2S(\op{b}+\hc{b})-\sqrt{\frac{S}{2}}\frac{\hc{b}\op{b}\op{b}+\hc{b}\hc{b}\op{b}}{2}\\
&-\sqrt{\frac{S}{2}}(\hc{b}\op{b}\op{b}+\hc{b}\op{b}\hc{b}+\op{b}\hc{b}\op{b}+\hc{b}\hc{b}\op{b})
\end{split}
\end{equation}
and Eq.~\eqref{eq:Hamiltonian_theta_SmSn} becomes
\begin{equation}
    \label{eq:Hamiltonian_theta_HP_detail}
    \begin{split}
    \op{H}_\mathrm{\theta} =& \hphantom{-}\gamma{}B_\mathrm{ext}\cos(\theta_{\rm{M}}-\theta)\hc{b}\op{b}\\
    &+\gamma{}B_\mathrm{ext}\sin(\theta_{\rm{M}}-\theta)\sqrt{\frac{S}{2}}\left[\op{b}+\hc{b}-\frac{\hc{b}\op{b}\op{b}}{4S}-\frac{\hc{b}\hc{b}\op{b}}{4S}\right]\\
    &+\sqrt{2S}\mathcal{C}\cos\theta_{\rm{M}}\sin\theta_{\rm{M}}\left[2S(\op{b}+\hc{b})-\frac{\hc{b}\op{b}\op{b}+\hc{b}\hc{b}\op{b}}{2}\right.\\
&-\left.(\hc{b}\op{b}\op{b}+\hc{b}\op{b}\hc{b}+\op{b}\hc{b}\op{b}+\hc{b}\hc{b}\op{b}))\right]\\
    &-2\mathcal{C}\cos^2\theta_{\rm{M}}\op{S}_m^2-2\mathcal{C}\sin^2\theta_{\rm{M}}\op{S}_n^2.
    \end{split}
\end{equation}
Adding the constant $\mathcal{C}\sin^2\theta_{\rm{M}}(\op{S}_n^2+\op{S}_y^2+\op{S}_m^2)$ in previous Eq.~\eqref{eq:Hamiltonian_theta_HP_detail} leads to
\begin{equation}
    \label{eq:Hamiltonian_theta_HP_added}
    \begin{split}
    \op{H}_\mathrm{\theta} =& \hphantom{-}\gamma{}B_\mathrm{ext}\cos(\theta_{\rm{M}}-\theta)\hc{b}\op{b}\\
    &+\gamma{}B_\mathrm{ext}\sin(\theta_{\rm{M}}-\theta)\sqrt{\frac{S}{2}}\left[\op{b}+\hc{b}-\frac{\hc{b}\op{b}\op{b}}{4S}-\frac{\hc{b}\hc{b}\op{b}}{4S}\right]\\
    &+\sqrt{2S}\mathcal{C}\cos\theta_{\rm{M}}\sin\theta_{\rm{M}}\left[2S(\op{b}+\hc{b})-\frac{\hc{b}\op{b}\op{b}+\hc{b}\hc{b}\op{b}}{2}\right.\\
&-\left.(\hc{b}\op{b}\op{b}+\hc{b}\op{b}\hc{b}+\op{b}\hc{b}\op{b}+\hc{b}\hc{b}\op{b}))\right]\\
    &-\mathcal{C}(2\cos^2\theta_{\rm{M}}-\sin^2\theta_{\rm{M}})\op{S}_m^2-\mathcal{C}\sin^2\theta_{\rm{M}}(\op{S}_n^2-\op{S}_y^2).
    \end{split}
\end{equation}
The terms are sorted by their order in
\begin{align}
    \label{eq:Hamiltonian_theta_HP_develop_1}
    \begin{split}
    \op{H}_{\mathrm{\theta}}^{(1)} =&\left[\gamma{}B_\mathrm{ext}\sin(\theta_{\rm{M}}-\theta)\sqrt{\frac{S}{2}}\right.\\
    &\left.+(2S)^{3/2}\mathcal{C}\cos\theta_{\rm{M}}\sin\theta_{\rm{M}}\vphantom{\sqrt{\frac{S}{2}}}\right](\op{b}+\hc{b})
    \end{split}\\
    \label{eq:Hamiltonian_theta_HP_develop_2}
    \begin{split}
    \op{H}_{\mathrm{\theta}}^{(2)}=&\left[\gamma{}B_\mathrm{ext}\cos(\theta_{\rm{M}}-\theta)+4\mathcal{C}S(\cos^2\theta_{\rm{M}}-\sin^2\theta_{\rm{M}}/2)\right]\hc{b}\op{b}\\
    &-\mathcal{C}S\sin^2\theta_{\rm{M}}(\op{b}\op{b}+\hc{b}\hc{b})
    \end{split}\\
    \label{eq:Hamiltonian_theta_HP_develop_3}
    \begin{split}
    \op{H}_{\mathrm{\theta}}^{(3)}=&-\gamma{}B_\mathrm{ext}\sin(\theta_{\rm{M}}-\theta)\frac{1}{4\sqrt{2S}}(\hc{b}\op{b}\op{b}+\hc{b}\hc{b}\op{b})\\
    &-\sqrt{\frac{S}{2}}\mathcal{C}\cos\theta_{\rm{M}}\sin\theta_{\rm{M}}(3\hc{b}\op{b}\op{b}+3\hc{b}\hc{b}\op{b}\\&+2\hc{b}\op{b}\hc{b}+2\op{b}\hc{b}\op{b})
    \end{split}\\
    \label{eq:Hamiltonian_theta_HP_develop_4}
    \begin{split}
    \op{H}_{\theta}^{(\rm{4})}=&\hphantom{+}\mathcal{C}\sin^2\theta_{\rm{M}}\frac{\hc{b}\op{b}\op{b}\op{b}+\op{b}\hc{b}\op{b}\op{b}+\hc{b}\hc{b}\op{b}\hc{b}+\hc{b}\hc{b}\hc{b}\op{b}}{4}\\
    &-2\mathcal{C}(\cos^2\theta_{\rm{M}}-\sin^2\theta_{\rm{M}}/2)(\hc{b}\op{b}\hc{b}\op{b}).
    \end{split}
\end{align}

With the static equilibrium defined as $\gamma{}B_{\rm{ext}}\sin(\theta_{\rm{M}}-\theta)=-4\mathcal{C}S\cos\theta_{\rm{M}}\sin\theta_{\rm{M}}$, the term of first order in $\hc{b}+\op{b}$ cancels, and the terms of third order simplify. For quartic terms, substituting $\op{b},\hc{b}$ by $\op{m},\hc{m}$ and only keeping terms of the correct order in the rotating wave approximation leads to
\begin{equation}
    \label{eq:Hamiltonian_theta_Bog_anharmfilt}
    \begin{split}
    \op{H}_{\theta}^{(\rm{4})}=\hbar^2\mathcal{C}&\left[
    \left(\alpha_1u^2v^2-\alpha_2u^3v\right)\hc{m}\hc{m}\op{m}\op{m}\right.\\
    &+\left(\alpha_1u^4-\alpha_2uv(u^2+v^2)/2\right)\hc{m}\op{m}\hc{m}\op{m}\\
    &+\left(\alpha_1u^2v^2-\alpha_2uv(u^2+v^2)/2\right)\hc{m}\op{m}\op{m}\hc{m}\\
    &+\left(\alpha_1u^2v^2-\alpha_2uv(u^2+v^2)/2\right)\op{m}\hc{m}\hc{m}\op{m}\\
    &+\left(\alpha_1v^4-\alpha_2uv(u^2+v^2)/2\right)\op{m}\hc{m}\op{m}\hc{m}\\
    &+\left.\left(\alpha_1u^2v^2-\alpha_2uv^3\right)\op{m}\op{m}\hc{m}\hc{m}
    \right],
    \end{split}
\end{equation}
where $\alpha_1=(\sin^2\theta_{\rm{M}}-2\cos^2\theta_{\rm{M}})$ and $\alpha_2=\sin^2\theta_{\rm{M}}$, which can be reordered to obtain \cref{eq:Hamiltonian_theta_Bog_short}.

\end{document}